\documentclass[aps,pra,reprint,amsmath,amssymb,longbibliography,floatfix,superscriptaddress,showkeys]{revtex4-2}
\usepackage[T1]{fontenc}
\usepackage{lmodern}
\usepackage{graphicx}
\usepackage{bm}
\usepackage{booktabs}
\usepackage{xcolor}
\usepackage{amsthm}
\usepackage{hyperref}
\graphicspath{{figures/}}
\newcommand{\rvec}{\mathbf r}
\newcommand{\dd}{\mathrm{d}}
\newcommand{\vA}{\mathbf A}
\newcommand{\eab}{e_{\rm Ab}}
\newcommand{\bsh}{b_{\rm sh}}
\newcommand{\brad}{b_{\rm rad}}
\newcommand{\btheta}{b_{\theta}}
\newcommand{\Y}{\mathcal Y}
\newcommand{\deltaZeroNineteen}{0.04068}

\newcommand{\etaNearNineteen}{4.94}
\newcommand{\bNearNineteen}{1.2930}
\newcommand{\bZeroNineteen}{1.1335}

\newcommand{\fitRootNineteen}{-12.57191}

\newcommand{\deltaZeroThirtySeven}{0.04146}

\newcommand{\etaNearThirtySeven}{1.79}
\newcommand{\bNearThirtySeven}{1.2724}
\newcommand{\bZeroThirtySeven}{1.1400}

\newcommand{\fitRootThirtySeven}{-12.56858}

\newcommand{\deltaZeroSixtyOne}{0.03487}

\newcommand{\etaNearSixtyOne}{1.52}
\newcommand{\bNearSixtyOne}{1.2552}
\newcommand{\bZeroSixtyOne}{1.1407}
\newcommand{\resZeroSixtyOne}{5.9\times10^{-8}}
\newcommand{\fitRootSixtyOne}{-12.57245}

\newcommand{\outerDefectPercent}{34.4}
\newcommand{\outerKineticPercent}{3.9}
\newcommand{\outerAmplitudePercent}{4.2}
\newcommand{\regionalLeakageRatio}{13.0}
\newcommand{\nearestDefectPercent}{45.8}
\newcommand{\nearestLeakageRatio}{19.7}

\newcommand{\responseMaxNineteen}{0.13}
\newcommand{\responseMaxThirtySeven}{0.78}
\newcommand{\responseMaxSixtyOne}{1.03}

\newcommand{\spatialrows}{%
-0.993$^\dagger$ & 45.8 & 4.1 & 4.2 & 2.4 & 19.7\\
-0.990$^\dagger$ & 42.9 & 4.1 & 4.2 & 2.4 & 17.6\\
-0.987$^\dagger$ & 36.8 & 3.9 & 4.1 & 2.4 & 14.2\\
-0.981$^\dagger$ & 35.0 & 3.9 & 4.1 & 2.3 & 13.4\\
-0.975 & 34.4 & 3.9 & 4.2 & 2.4 & 13.0\\
-0.969 & 33.5 & 3.9 & 4.2 & 2.4 & 12.5\\
-0.957 & 31.5 & 3.9 & 4.2 & 2.4 & 11.4\\
-0.945 & 29.7 & 3.9 & 4.3 & 2.4 & 10.3\\
-0.886 & 23.4 & 4.0 & 4.6 & 2.4 & 7.3\\
-0.827 & 19.9 & 4.1 & 4.8 & 2.5 & 5.8\\
-0.709 & 15.5 & 4.2 & 5.2 & 2.6 & 4.1\\
-0.473 & 10.8 & 4.4 & 5.8 & 2.7 & 2.6\\
-0.236 & 8.5 & 4.5 & 6.3 & 2.8 & 2.0\\
0.000 & 4.6 & 6.1 & 7.1 & 3.0 & 0.7\\
}

\newcommand{\diagnosticrows}{%
19 & -0.993 & 0.8250 & $3.7\times10^{-4}$ & $6.5\times10^{-4}$ & $1.1\times10^{-4}$\\
19 & 0.000 & 8.9271 & 0.0407 & $3.0\times10^{-8}$ & $7.9\times10^{-8}$\\
37 & -0.993 & 1.1248 & $1.3\times10^{-4}$ & $1.4\times10^{-4}$ & $1.9\times10^{-4}$\\
37 & 0.000 & 12.4977 & 0.0415 & $6.8\times10^{-4}$ & $7.8\times10^{-4}$\\
61 & -0.993 & 1.4454 & $1.2\times10^{-4}$ & $2.3\times10^{-5}$ & $9.3\times10^{-3}$$^\dagger$\\
61 & -0.975 & 2.6364 & $3.8\times10^{-4}$ & $2.1\times10^{-6}$ & $4.1\times10^{-4}$\\
61 & 0.000 & 16.0007 & 0.0349 & $3.4\times10^{-8}$ & $5.9\times10^{-8}$\\
}

\newcommand{\confirmrows}{%
continuation, start & 2.636447 & $3.8\times10^{-4}$ & $4.1\times10^{-4}$ & 34.4 & 13.0 & 53\\
+5$\times$2000 iterations & 2.636437 & $3.8\times10^{-4}$ & $3.5\times10^{-4}$ & 34.3 & 12.9 & 53\\
grid $N_{\rm g}=800$ & 2.636435 & $3.8\times10^{-4}$ & $2.5\times10^{-4}$ & 34.2 & 12.8 & 53\\
box $1.3L$ & 2.636433 & $3.8\times10^{-4}$ & $2.4\times10^{-4}$ & 34.2 & 12.9 & 53\\
fresh start lattice a & 2.642155 & $2.8\times10^{-4}$ & $7.7\times10^{-3}$$^\dagger$ & 46.4 & 17.6 & 56\\
fresh start lattice b & 2.669774 & $3.6\times10^{-4}$ & $7.8\times10^{-3}$$^\dagger$ & 32.6 & 10.8 & 55\\
fresh start random a & 4.402882 & $6.5\times10^{-3}$ & $6.7\times10^{-1}$$^\dagger$ & 39.5 & 8.4 & 58\\
}

\newcommand{\confirmzerorows}{%
polish & 16.000723 & 0.03487 & $5.9\times10^{-8}$ & 4.6 & 51\\
lattice & 16.009972 & 0.03498 & $2.8\times10^{-4}$ & 5.7 & 52\\
random & 16.158620 & 0.05091 & $4.0\times10^{-3}$$^\dagger$ & 7.9 & 54\\
}

\begin{document}
\title{Near self-duality and spatial reorganization of anyon vortex lattices}
\author{Bao-Duy Le}
\affiliation{Faculty of Physics and Engineering Physics, University of Science, Ho Chi Minh City, Vietnam}
\affiliation{Vietnam National University, Ho Chi Minh City, Vietnam}
\email[Email: ]{24130012@student.hcmus.edu.vn}
\begin{abstract}
Contact attraction drives the self-generated vortex lattice of a trapped anyon gas towards self-duality, but integrated energies do not reveal where the remaining departure from it is located. We address this question in the almost-bosonic Chern--Simons--Schr\"odinger average-field model using a positive, gauge-invariant local defect that measures the covariant kinetic energy in excess of the magnetic bound. Numerically computed low-energy states at three statistical fluxes show that, as the stability line is approached, the integrated defect falls by about two orders of magnitude while the density modulation grows, so that magnetic saturation and density uniformity separate. The residual defect concentrates in a vortex-free outer shell that carries unscreened statistical flux. There it reduces to a smooth radial mismatch between the statistical flow and the density decay, which the interior lattice cannot cancel, whereas the covariant kinetic and amplitude-gradient energies show no comparable enrichment. The effect is robust to grid and box refinement, while its magnitude depends on flux and on the low-energy branch. Exact dilation and contact-response identities relate the squared energy to the density shape and provide consistency checks. Repulsion produces the opposite trend: a smoother density with a larger defect.
\end{abstract}
\keywords{Anyons; Chern--Simons--Schr\"odinger theory; self-duality; vortex lattices; nonlinear Landau levels; average-field theory; attractive interactions}
\maketitle
\section{Introduction}\label{sec:intro}
Fractional exchange statistics permits quantum particles in two dimensions to interpolate beyond the Bose and Fermi alternatives \cite{LeinaasMyrheim1977,Wilczek1982}. Fractional quantum Hall quasiparticles provide a prominent realization \cite{Arovas1984}, with collision and interference experiments revealing signatures of fractional statistics \cite{Bartolomei2020,Nakamura2020}. The collective behavior of a gas of many mobile anyons is a further problem: statistics, spatial organization, and interactions jointly determine its energy and stability. An overview of the available many-anyon descriptions is given in Ref.~\cite{LundholmReview2024}.

In the magnetic-gauge representation, statistical phases are encoded by particle-attached fluxes. Close to the bosonic limit, the collective field of many weak attachments leads to a self-consistent average-field functional. Its derivation for extended anyons and its local-density regime have been established in Refs.~\cite{LundholmRougerie2015,CorreggiLundholmRougerie2017}, with more recent progress on dilute regimes \cite{Visconti2026}. A concrete large-flux prediction is a lattice of counter-rotating vortices, whose local spacing follows the particle density \cite{CorreggiDuboscqLundholmRougerie2019}.

Attraction introduces a competition between statistical kinetic pressure and concentration. In the ordinary two-dimensional focusing nonlinear Schr\"odinger model, the common scaling of kinetic and quartic energies leads to the Townes threshold and sharp interpolation inequalities \cite{ChiaoGarmireTownes1964,Weinstein1983}. The self-generated magnetic field changes the sharp inequality and its saturating states. Self-dual Chern--Simons solitons were constructed by Jackiw and Pi \cite{JackiwPi1990,JackiwPi1990PRD}; their relation to magnetic stability and nonlinear Landau levels (NLLs) has since been established mathematically \cite{AtaeiLundholmNguyen2024}. Microscopic derivations and trapped CSS spectra, including attractive and repulsive interactions and NLL trial states, are developed in Refs.~\cite{AtaeiGirardotLundholm2025,AtaeiEtAl2025}.

Those results establish the stability line and the exact self-dual family. The question addressed here is spatial: how is the departure from self-duality distributed across a many-vortex cloud as attraction is increased? A small excess above the magnetic kinetic bound need not be distributed like particle mass, kinetic energy, or density gradients. Identifying that distribution gives information that integrated spectra and collapse thresholds leave unresolved.

We study this question through a positive gauge-invariant defect, a density-shape response, and regional measures of magnetic nonsaturation. The relative integrated defect decreases by about two orders of magnitude on attractive branches at $\beta=19,37,61$, while density modulation increases. At $\beta=61$, reconstruction over the full attractive sequence reveals progressive peripheral enrichment in the strong-attraction regime. In the closest state satisfying our stationarity criterion, the outer tenth of particle mass carries $\outerDefectPercent\%$ of the defect but only $\outerKineticPercent\%$ of the covariant kinetic energy and $\outerAmplitudePercent\%$ of the amplitude-gradient energy; less converged states still closer to the boundary continue the trend. The same enrichment occurs at $\beta=19$, and the phase winding shows that the outermost fifth of the mass contains no vortices, so a few statistical flux units remain unscreened there. The effect therefore resolves a spatial imbalance in the cancellation that produces near self-duality, beyond the distribution of the positive gradient cost alone.

An exact separation of amplitude gradients, gauge-invariant flow, and their signed cross term provides a local interpretation. A complementary dilation argument connects the squared energy to density shape and gives an integral contact-response check using the computed branches. These identities organize the observations; the new numerical content is the interaction-dependent spatial redistribution and its comparison with independent local energy baselines.

The distinction from prescribed-field vortex lattices is useful. Abrikosov and lowest-Landau-level theories characterize cell-averaged density factors in an external magnetic or rotational environment \cite{Abrikosov1957,KleinerRothAutler1964,CooperKomineasRead2004,AftalionBlancDalibard2005}. Here the magnetic field follows the density. We distinguish the global trapped shape factor from a homogeneous cell factor and resolve radial and angular contributions. Repulsion supplies a complementary comparison: density becomes smoother while magnetic saturation deteriorates.

Sections~\ref{sec:model}--\ref{sec:method} specify the model, response identities, and numerical observables. Sections~\ref{sec:attraction} and \ref{sec:spatial} present the attractive response and spatial mechanism; Sec.~\ref{sec:repulsion} treats repulsion. Variational interpretation and the scope of the conclusions are discussed in Secs.~\ref{sec:variational} and \ref{sec:discussion}. Analytic details and numerical diagnostics are given in the appendices and Supplemental Material (SM).

\section{Average-field model and stability}\label{sec:model}
\subsection{Functional and coupling conventions}
We use the planar coordinate $\rvec=(x,y)\in\mathbb R^2$ and the radial distance $r=|\rvec|$. The normalized Chern--Simons--Schr\"odinger (CSS) energy is
\begin{equation}
 \begin{split}
 \mathcal E_{\beta,G}[u]=\int_{\mathbb R^2}\bigg[&|(-\mathrm{i}\nabla+\beta\vA[\rho])u|^2\\
 &+|\rvec|^2\rho+\frac G2\rho^2\bigg]\,\dd\rvec,\\
 &\rho=|u|^2,\qquad\int_{\mathbb R^2}\rho(\rvec)\,\dd\rvec=1,
 \end{split}\label{eq:functional}
\end{equation}
in units $\hbar=2m=1$. The self-consistent potential is
\begin{equation}
 \vA[\rho]=\nabla^\perp(\log|\cdot|*\rho),\qquad
 \operatorname{curl}(\beta\vA)=2\pi\beta\rho,
 \label{eq:gauge}
\end{equation}
where $\nabla^\perp=(-\partial_y,\partial_x)$ and $\dd\rvec$ denotes two-dimensional Lebesgue measure. Thus $\beta$ is the total number of statistical flux units for unit mass. We take $\beta>0$; complex conjugation maps the opposite sign to this convention. The sign of $G$ distinguishes focusing attraction from defocusing repulsion. We write
\begin{equation}
 \begin{gathered}
 T=\int_{\mathbb R^2}|\bm D u|^2\,\dd\rvec,\quad
 I=\int_{\mathbb R^2}\rho^2\,\dd\rvec,\\
 P=\int_{\mathbb R^2}|\rvec|^2\rho\,\dd\rvec,\quad
 \bm D=-\mathrm{i}\nabla+\beta\vA[\rho],
 \end{gathered}
 \label{eq:parts}
\end{equation}
so that the energy $E=\mathcal E_{\beta,G}[u]$ equals $T+GI/2+P$ for any field. The admissible continuum class is normalized $H^1(\mathbb R^2)$ fields with finite second moment and finite magnetic energy.

The parameter $G$ is an effective scalar coupling. It should not be identified without further analysis with a bare zero-range interaction at fixed nonzero exchange statistics. The latter involves a choice of self-adjoint realization even in the two-anyon problem \cite{CorreggiOddis2018}. The microscopic CSS construction distinguishes contributions from spin-orbit coupling and the flux-profile scale \cite{AtaeiGirardotLundholm2025}. The effective attractive functional is therefore a meaningful variational model independently of whether a particular microscopic regularization has been proved to converge to it.

The usual almost-bosonic scaling first takes particle number $N\to\infty$ with exchange parameter $\alpha\to0$ and $\alpha N\simeq\beta$ fixed. Studying increasing $\beta$ is a subsequent analysis of that effective functional; it is not a proof of a simultaneous many-body limit. Results on weak-field attractive collapse \cite{Nguyen2024} concern a different scaling from the dense-vortex branches considered here. Fermionic-gauge Hartree and magnetic Thomas--Fermi descriptions \cite{LevittLundholmRougerie2026} and rigorous exclusion estimates for ideal anyons \cite{LundholmSeiringer2018} address related but distinct limits. In particular, the coefficient of the present magnetic inequality is not an assumption of fermionic occupancy.

\subsection{Known threshold and self-dual states}
For Eq.~\eqref{eq:gauge}, the Bogomolny identity is
\begin{equation}
 \begin{gathered}
 T=2\pi\beta I+Q,\qquad Q=\int_{\mathbb R^2}d(\rvec)\,\dd\rvec\geq0,\\
 d(\rvec)=|(D_x-\mathrm{i}D_y)u|^2.
 \end{gathered}
 \label{eq:bogo}
\end{equation}
The opposite covariant combination has integral $T+2\pi\beta I$. The sign in Eq.~\eqref{eq:bogo} therefore matters for spatial interpretation. The identity follows from $[D_x,D_y]=-\mathrm{i}\,2\pi\beta\rho$ and integration by parts. Although the phase of $u$ is gauge dependent, $d$ is gauge invariant.

Let $C_*(\beta)$ be the infimum of $T/I$ over normalized admissible fields. The sharp result is $C_*(\beta)=2\pi\beta$ for $\beta\geq2$, the regime relevant here; the self-dual (supersymmetric) structure behind it is lost at weaker coupling \cite{AtaeiLundholmNguyen2024,Lundholm2024}. Consequently the stability boundary of Eq.~\eqref{eq:functional} is
\begin{equation}
 G_c=-4\pi\beta,\qquad \beta\geq2.
 \label{eq:critical}
\end{equation}
For $G<G_c$, a field with $T+GI/2<0$ can be concentrated by dilation, sending its energy to $-\infty$. For $G\geq G_c$, the nontrapped part is nonnegative. This statement concerns boundedness of the infimum; it does not specify attainment of the infimum at the boundary or the limiting trapped energy along a given flux sequence.

Regular normalized fields with $Q=0$ exist only at even flux $\beta=2n$, $n\in\mathbb N$ \cite{AtaeiLundholmNguyen2024,Lundholm2024}. They are NLL states associated with rational polynomial data, building on the self-dual Chern--Simons solitons \cite{JackiwPi1990,JackiwPi1990PRD}. The principal numerical fluxes $19,37,61$ are all odd. Their energetic proximity to saturation can be quantified, but it cannot be equated with convergence to a regular exact NLL at the same fixed flux without an additional limiting argument.

\subsection{Distance to the stability boundary}
We use the dimensionless variables
\begin{equation}
 \kappa=\frac{G}{4\pi\beta},\qquad
 \epsilon=1+\kappa,\qquad a=2\pi\beta\epsilon,
 \label{eq:couplings}
\end{equation}
so that attraction occupies $-1<\kappa<0$ and the established boundary is $\epsilon=0$. The source calculations use an additional label,
\begin{equation}
 \begin{gathered}
 \gamma=\frac{G}{2\eab\beta},\qquad \kappa=\frac{\eab}{2\pi}\gamma,\\
 \eab=7.424437329108944.
 \end{gathered}
 \label{eq:gamma}
\end{equation}
The source calculations fix $\eab$ to the value in Eq.~\eqref{eq:gamma} as a zero-contact lattice reference; the corresponding homogeneous coefficient was studied in Ref.~\cite{CorreggiDuboscqLundholmRougerie2019}. Here it only converts coupling labels; all results below are quoted in $\kappa$ or $\epsilon$, and the $\gamma$ labels appear only in the Supplemental Material. The closest sampled coupling, $\gamma=-0.84$, corresponds to $\kappa=-0.992574$ and $\epsilon=0.00742584$.

\begin{figure*}[t]
\centering
\includegraphics[width=\textwidth]{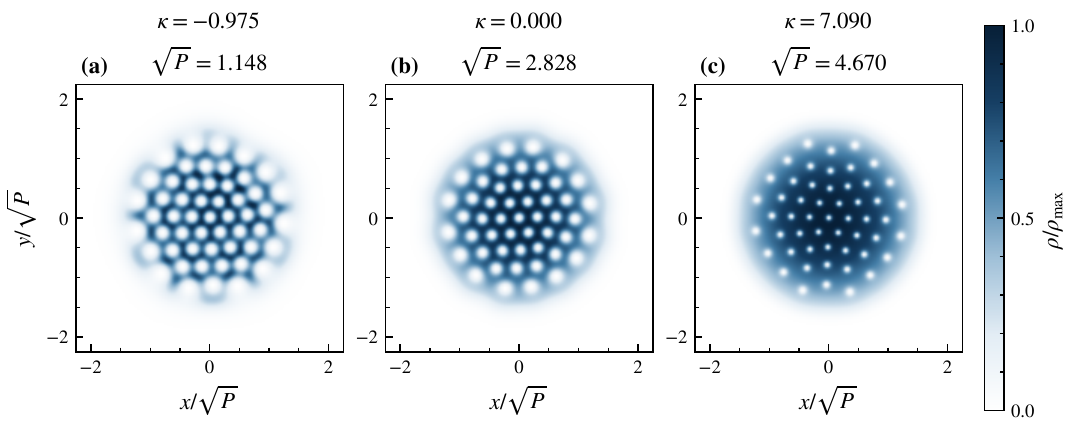}
\caption{Density across the interaction axis at $\beta=61$: strong attraction, zero contact coupling, and repulsion, from left to right. Coordinates are divided by the root-mean-square radius $\sqrt P$; the displayed radii are in trap units. Each panel uses its own $\rho_{\max}$ on a common normalized color scale. A resolved vortex array persists through substantial physical compression, while repulsion weakens relative density variation outside the cores. Display interpolation does not enter any integral or diagnostic.}
\label{fig:density}
\end{figure*}

\section{Self-duality, density shape, and energy response}\label{sec:theory}
\subsection{An exact separation of defect, shape, and scale}
Define the relative self-duality defect by
\begin{equation}
 \delta=\frac{Q}{2\pi\beta I}=\frac{T}{2\pi\beta I}-1\geq0.
 \label{eq:delta}
\end{equation}
It measures the excess covariant kinetic energy relative to the magnetic quartic scale of the same density. Its smallness is an energetic statement. It does not imply uniform density, disappearance of vortices, or norm convergence to an exact NLL. Exactly,
\begin{equation}
 E=Q+aI+P=2\pi\beta(\epsilon+\delta)I+P.
 \label{eq:decomp}
\end{equation}
Two scale-invariant quantities are useful:
\begin{equation}
 \bsh=\frac{9\pi}{4}IP,\qquad
 \eta=\frac{Q}{aI}=\frac{\delta}{\epsilon}.
 \label{eq:shapedef}
\end{equation}
The first characterizes density structure; the second compares the defect with the remaining attractive stiffness. Since $a\to0$ at the boundary, small $\delta$ need not imply negligible $\eta$.

For unit mass, the density inequality
\begin{equation}
 IP\geq\frac{4}{9\pi}
 \label{eq:densitybound}
\end{equation}
follows by minimizing $I$ at fixed second moment. Its equality density is an inverted parabola (Appendix~\ref{app:density}). Thus $\bsh\geq1$ independently of any vortex or Landau-level approximation. The full density enters $I$, so $\bsh$ includes the envelope and microscopic modulation. It is not a homogeneous Abrikosov cell factor.

A finite-resolution field need not satisfy virial stationarity. We retain
\begin{equation}
 w=\frac{Q+aI-P}{P}=\frac{E-2P}{P}.
 \label{eq:virialdef}
\end{equation}
For $a>0$, direct algebra gives
\begin{equation}
 E^2=\frac{32\beta}{9}\bsh(\epsilon+\delta)\mathcal V(w),
 \qquad\mathcal V(w)=1+\frac{w^2}{4(1+w)}.
 \label{eq:shapeidentity}
\end{equation}
This identity applies to every admissible field, not only stationary solutions. Since $Q+aI>0$, one has $w>-1$. The virial correction is quadratic near $w=0$; this does not control errors in directions orthogonal to dilation. A good virial check can coexist with a substantial stationary residual.

The dimensionless squared energy
\begin{equation}
 \Y(\kappa)=\frac{9E^2}{32\beta}
 =\bsh(\epsilon+\delta)\mathcal V(w)
 \label{eq:Y}
\end{equation}
makes the separate contributions transparent. The local magnetic scale supplies $\epsilon$, the failure of saturation adds $\delta$, and the density geometry multiplies their sum. In particular, attraction can suppress $\delta$ while increasing $\bsh$. These changes are not contradictory: they refer to different constraints on the field.

\subsection{Dilation and the geometry of squared energy}
The mass-preserving transformation $u_\lambda(\rvec)=\lambda u(\lambda\rvec)$ gives
\begin{equation}
 T_\lambda=\lambda^2T,\qquad I_\lambda=\lambda^2I,\qquad
 P_\lambda=\lambda^{-2}P.
 \label{eq:scaling}
\end{equation}
The statistical potential has the same inverse-length scaling as the gradient. The additive constant generated by scaling the logarithmic potential disappears after applying $\nabla^\perp$. With $\mathcal T_G=T+GI/2=Q+aI$,
\begin{equation}
 E[u_\lambda]=\lambda^2\mathcal T_G+\lambda^{-2}P.
 \label{eq:dilationenergy}
\end{equation}
Stationarity against this variation yields $\mathcal T_G=P$, hence $w=0$ and $E=2P$. For a trial shape with $\mathcal T_G>0$, the best scale satisfies $\lambda_*^4=P/\mathcal T_G$ and has energy $2\sqrt{\mathcal T_GP}$.

Let $E_{\rm gs}(G)$ be the ground-state infimum at fixed $\beta$. The admissible class is invariant under dilation, and its energies are positive for $a>0$. Optimizing over scale and then over shapes gives
\begin{equation}
 \begin{gathered}
 E_{\rm gs}(G)^2=\inf_{\|u\|_2=1,\,P<\infty}
 4P[u]\left(T[u]+\frac G2I[u]\right),\\
 a>0.
 \end{gathered}
 \label{eq:concavity}
\end{equation}
Each shape supplies an affine function of $G$ to the squared energy, so their lower envelope is concave. This strengthens the direct variational concavity of $E_{\rm gs}$ itself. Squaring an arbitrary positive concave function would not establish the result; the dilation representation is the key step.

Equation~\eqref{eq:concavity} also distinguishes scale relaxation from shape changes. If one shape stays approximately optimal over an interval, $E^2$ will be approximately affine there. Curvature indicates deformation or selection among shapes, even when each field has a small magnetic defect. A finite set of stored shapes realizes a piecewise-affine version of this construction without new minimization (Sec.~\ref{sec:variational}).

Combining the magnetic and density bounds yields
\begin{equation}
 E_{\rm gs}(G)^2\geq\frac{16a}{9\pi}
 =\frac{32\beta}{9}\epsilon.
 \label{eq:lower}
\end{equation}
For any fixed exact NLL $v$ with finite $I[v]$ and $P[v]$, the analogous trial bound is $E_{\rm gs}^2\leq4aI[v]P[v]$. The finite-moment condition is essential: generic rational NLL densities can have an $r^{-4}$ tail and logarithmically divergent $P$. The existing NLL variational approach \cite{AtaeiEtAl2025} supplies useful upper bounds when the tail condition is satisfied or treated separately. A vanishing lower bound alone does not prove vanishing trapped energy.

\subsection{Contact derivatives and fitted slopes}
On a differentiable stationary branch, Hellmann--Feynman gives
\begin{equation}
 \frac{\dd E}{\dd G}=\frac I2.
 \label{eq:HF}
\end{equation}
Using $E=2P$ at a virial stationary point,
\begin{equation}
 \frac{\dd E^2}{\dd G}=EI=2PI=\frac{8}{9\pi}\bsh,
 \qquad \frac{\dd\Y}{\dd\kappa}=\bsh.
 \label{eq:E2slope}
\end{equation}
The squared-energy slope therefore measures density shape even away from self-duality. Ground-state concavity implies that this response is nonincreasing with increasing $\kappa$, where the derivative exists. A smooth metastable branch can obey Eq.~\eqref{eq:E2slope} without satisfying the global concavity constraint. At a crossing, one-sided slopes replace a unique derivative.

For a local fit $E^2=\sigma_{\rm fit}(G-G_{\rm int})$, the coefficient $b_{\rm fit}=9\pi\sigma_{\rm fit}/8$ estimates a shape response. Interpreting it as a limiting critical constant requires further control of shape variation and $\eta$. In particular, the approximation $E^2\simeq16a\bsh/(9\pi)$, which by Eqs.~\eqref{eq:shapedef} and \eqref{eq:shapeidentity} requires $\eta\ll1$ and $w\simeq0$, does not follow from $\delta\ll1$ alone when $\epsilon$ is also small.

\subsection{Radial and angular density structure}
For the radial trap, define
\begin{equation}
 \overline\rho(r)=\frac{1}{2\pi}\int_0^{2\pi}\rho(r,\theta)\,\dd\theta,
 \quad I_{\rm rad}=2\pi\int_0^\infty r\overline\rho(r)^2\,\dd r.
 \label{eq:radialaverage}
\end{equation}
Angular averaging preserves mass and $P$. Orthogonality of the mean and fluctuations gives
\begin{equation}
 I=I_{\rm rad}+I_{\rm var},\qquad
 I_{\rm var}=\int_{\mathbb R^2}[\rho(\rvec)-\overline\rho(|\rvec|)]^2\,\dd\rvec\geq0.
 \label{eq:variance}
\end{equation}
Consequently,
\begin{equation}
 \bsh=\brad\btheta,\qquad
 \brad=\frac{9\pi}{4}P I_{\rm rad},\qquad
 \btheta=\frac{I}{I_{\rm rad}}\geq1.
 \label{eq:radialfactors}
\end{equation}
In the continuum, $\brad\geq1$ by Eq.~\eqref{eq:densitybound}. This is a rotational decomposition about the trap center, not a unit-cell homogenization. Radial oscillations from vortex shells remain in $\overline\rho$, while $\btheta$ measures nonaxisymmetric modulation. It separates two sources of quartic cost without requiring a locally uniform cloud. Its annular discretization has a controlled bookkeeping identity and a measurable bin-width sensitivity, described below.

\section{Numerical representation and observables}\label{sec:method}
\subsection{Free-space field and constrained optimization}
The functional is discretized on a square grid using Fourier derivatives for $u$. The density is zero padded onto a threefold larger array and convolved with a truncated logarithmic kernel. This follows the free-space construction of Vico, Greengard, and Ferrando \cite{VicoGreengardFerrando2016}. Its cutoff exceeds the largest source--target separation within the physical square. The representation therefore does not replace $\rho$ in the magnetic field by its deviation from the cell average, as an unpadded periodic Poisson multiplier would. The precise multiplier and padding convention are given in Appendix~\ref{app:kernel}.

The field dependence of the kinetic energy must be differentiated through $\vA[\rho]$. With $\bm K=\nabla^\perp\log|\cdot|$ and $J_j=\operatorname{Re}(\bar uD_ju)$, one finds
\begin{equation}
 g:=\frac{\delta\mathcal E}{\delta\bar u}
 =\bm D^2u+|\rvec|^2u+G\rho u
 -2\beta u\sum_{j=x,y}K_j*J_j.
 \label{eq:gradient}
\end{equation}
The last term is the density backreaction. Omitting it would optimize a different energy even if the instantaneous vector potential were evaluated accurately.

The archived solver uses preconditioned nonlinear conjugate gradients on the normalization sphere, with an energy-decreasing line search. Normalized gradient-flow methods for nonlinear Bose fields provide the broader numerical setting \cite{BaoDu2004,BaoCai2013}. The $\beta=61$ refinement additionally optimizes the dilation mode at intervals, accepting a dilation only if it lowers the energy. Initial fields include density-mapped vortex arrays and random starts; continuation is used along the interaction axis. Comparing starts reduces, but does not eliminate, the possibility of trapping in local minima.

Stationarity is measured by
\begin{equation}
 r_{\rm rel}=\frac{\|g-\mu u\|_2}{\max(1,|\mu|)},\qquad
 \mu=\operatorname{Re}\langle u,g\rangle.
 \label{eq:residual}
\end{equation}
The optimizer stops at $r_{\rm rel}<10^{-7}$, which is reached away from the boundary but not by most near-critical states, where the dilation mode and lattice rearrangements become soft. We therefore adopt the practical acceptance criterion
\begin{equation}
 r_{\rm rel}\leq10^{-3}
 \label{eq:criterion}
\end{equation}
for every quantitative statement, fit, and representative map. States above it are retained in the record, drawn with open symbols, and marked by a dagger in tables; they are used only to indicate trends. Independent diagnostics include $w$, mass normalization, boundary density, and the Bogomolny identity. They test different features: a small energy drift need not imply a small gradient, while a small dilation derivative does not imply stationarity against lattice rearrangements.

\subsection{Branch selection and numerical precision}
The attractive comparison uses one complete specified branch per flux: 11 states each at $\beta=19,37$ from the long-run checkpoint, and 14 at $\beta=61$ from the refined checkpoint. Their grids are $384^2$, $512^2$, and $640^2$, with sides $L=13.81850$, $15.59688$, and $17.20544$. The first two sequences allow up to 50000 iterations. The $\beta=61$ sequence has a base budget of 5000 and an extended near-boundary budget, with earlier termination if the line search stalls. Sources are fixed at the branch level before post-processing. Criterion~\eqref{eq:criterion} is met by all 11 states at $\beta=19$, by 10 of 11 at $\beta=37$ (the exception is $\kappa=-0.945$, $r_{\rm rel}=1.3\times10^{-3}$), and by 10 of 14 at $\beta=61$; the $\beta=61$ exceptions are the four states with $\kappa\leq-0.981$ ($1.8\times10^{-3}\leq r_{\rm rel}\leq9.3\times10^{-3}$). The $\beta=61$ zero-coupling field was re-converged in a separate confirmation run (Sec.~\ref{sec:robust}) to $r_{\rm rel}=\resZeroSixtyOne$; its energy is $0.36\%$ below that of the original continuation start, and it replaces that start in all zero-coupling comparisons.

The repulsive comparison has five positive couplings at each flux. The $\beta=19,37$ states come from the separate free-space interaction scan on $352^2$ and $480^2$ grids; the $\beta=61$ states come from the refined calculation on $640^2$. Branch selection is fixed in advance and preserves each sequence's box and initialization history. All fifteen repulsive states satisfy Eq.~\eqref{eq:criterion}, with $r_{\rm rel}\leq4.1\times10^{-4}$; their boxes have sides $L=19.70$, $22.75$, and $26.76$, and the mass in the boundary cells is below $10^{-11}$.

The selected fields are therefore treated as low-energy numerical branches, with ground-state asymptotics discussed separately in Sec.~\ref{sec:discussion}. Representative stationary residuals, fit-window sensitivity, and response checks are supplied in the Supplemental Material (SM). Detailed computational records are retained in the separate reproducibility archive.

The free-space kernel is checked against the analytic vector potential of a normalized Gaussian. The functional derivative is checked against central finite differences, and the saturating sign is tested independently. These implementation tests complement the optimization diagnostics. All spatial observables are reconstructed directly from the complex field $u$.

\subsection{Spatial and annular observables}
The representative attractive $\beta=61$ maps use $\kappa=0,-0.709,-0.975$; the last is the closest state satisfying Eq.~\eqref{eq:criterion}. Their defect is reconstructed with the covariant sign in Eq.~\eqref{eq:bogo}, and its integral is compared with $T-2\pi\beta I$. Display interpolation is used only in the figures; normalization, energy integrals, and cumulative distributions use the original grid.

Radial averages are computed by averaging pixels in annuli of width $h=L/N_{\rm g}$, where $N_{\rm g}$ is the number of grid points per direction. Repeating with widths $2h$ and $4h$ measures the decomposition's sensitivity. The discrete radial-mean square plus angular variance must reproduce $I$ exactly. Broader annuli average out some radial shell structure and therefore transfer part of it into the measured variance; this is a resolution dependence, not a confidence interval from random sampling.

To locate the defect without choosing a cloud-support threshold, we use
\begin{equation}
 M(R)=\int_{B_R}\rho(\rvec)\,\dd\rvec,\qquad
 F_d(R)=\frac{1}{Q}\int_{B_R}d(\rvec)\,\dd\rvec.
 \label{eq:cumulants}
\end{equation}
Here $B_R=\{\rvec\in\mathbb R^2:|\rvec|\leq R\}$. The curve $F_d$ versus $M$ compares radial distributions. A diagonal curve corresponds to identical radial distributions of mass and defect. We also report the fraction $1-F_d(R_\nu)$ outside the radius $R_\nu$ containing mass fraction $\nu$, chiefly $R_{0.9}$. On the grid, $R_\nu$ is the first discrete radius reaching $\nu$ and the outer integral uses the strict mask $r>R_\nu$. This fixes the treatment of the pixel shell at the crossing.

\section{Attractive response: magnetic saturation and density shape}\label{sec:attraction}
\subsection{Integrated defect and the remaining energy scale}
Figure~\ref{fig:density} shows a resolved vortex array across substantial physical compression. At $\beta=61$, the root-mean-square radius decreases from $2.828$ at zero contact coupling to $1.148$ at $\kappa=-0.975$, while reaching $4.670$ under the strongest displayed repulsion. Scaling coordinates by $\sqrt P$ separates physical size from the internal density organization.

\begin{figure*}[t]
\centering
\includegraphics[width=\textwidth]{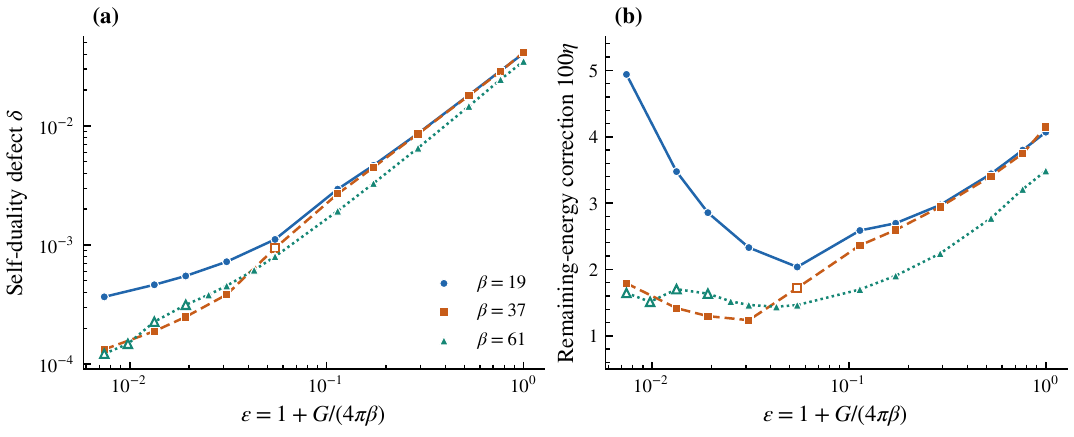}
\caption{Integrated attractive response. (a) Relative magnetic defect $\delta$ versus distance $\epsilon$ from the established stability line. (b) The defect relative to the remaining positive quartic coefficient, $\eta=\delta/\epsilon$, expressed as a percentage. Lines connect computed states; open symbols mark states above the stationarity criterion, Eq.~\eqref{eq:criterion}. A small magnetic defect can remain a percent-level correction to the residual energy near the boundary.}
\label{fig:defect}
\end{figure*}

At $G=0$, the selected $\beta=19,37,61$ states have $\delta=\deltaZeroNineteen,\deltaZeroThirtySeven,\deltaZeroSixtyOne$. At the closest accepted couplings these decrease to $3.67\times10^{-4}$ and $1.33\times10^{-4}$ at $\epsilon=0.0074$ for $\beta=19,37$, and to $3.82\times10^{-4}$ at $\epsilon=0.025$ for $\beta=61$ [Fig.~\ref{fig:defect}(a)]. The four closer $\beta=61$ states, above criterion~\eqref{eq:criterion}, continue the decrease to $1.22\times10^{-4}$ at $\epsilon=0.0074$. Because $\delta$ is invariant under dilation, this decrease reflects a change of shape and phase relative to the self-generated field, beyond compression alone.

At the closest accepted couplings, $\eta$ is $\etaNearNineteen\%$, $\etaNearThirtySeven\%$, and $\etaNearSixtyOne\%$ for increasing flux. The upturn for $\beta=19$ in the last points shows why $\delta\ll1$ alone is insufficient to infer an endpoint asymptotic. The two scales $\delta$ and $\epsilon$ must be compared.

The full-density shape factor changes in the opposite direction. Between zero coupling and the closest accepted point it rises from $\bZeroNineteen$ to $\bNearNineteen$ at $\beta=19$, from $\bZeroThirtySeven$ to $\bNearThirtySeven$ at $\beta=37$, and from $\bZeroSixtyOne$ to $\bNearSixtyOne$ at $\beta=61$. Increasing magnetic saturation therefore coexists with an increasing quartic cost of density structure. This separation is central to interpreting the lattice response.

At zero contact coupling, the exact relation becomes
\begin{equation}
 \frac{9\pi E^2}{16\beta}=2\pi\bsh(1+\delta)\mathcal V(w).
 \label{eq:effectivezero}
\end{equation}
Density shape and magnetic nonsaturation both enhance the trapped coefficient above $2\pi$. The thermodynamic reference $\eab$ can only be identified after the appropriate large-flux and optimization limits; the finite-flux relation already explains how the two contributions combine.

\subsection{Squared energy as a contact-response observable}
\begin{figure*}[t]
\centering
\includegraphics[width=\textwidth]{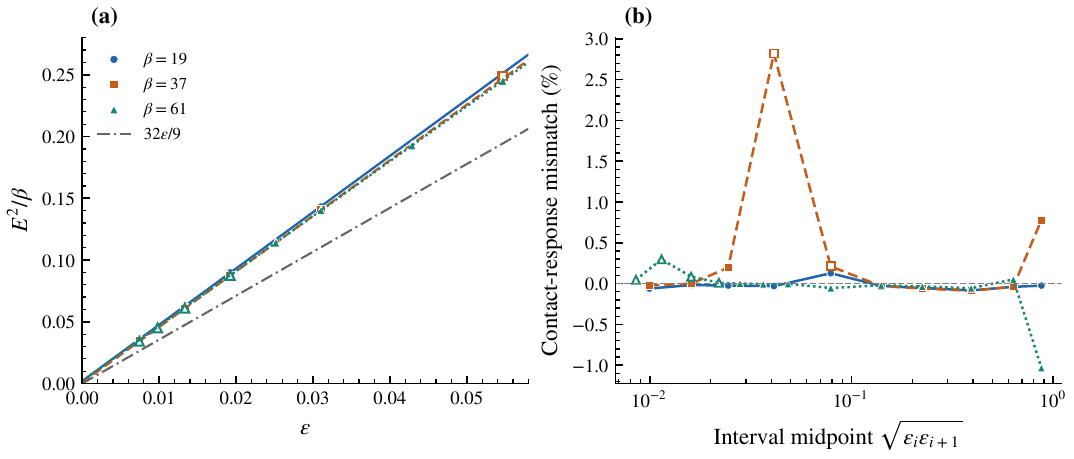}
\caption{Squared energy and contact-response consistency. (a) Near-boundary $E^2/\beta$ with fits to the three closest accepted points and the rigorous lower-bound reference. (b) Relative mismatch $100\mathcal R_i$ between the squared-energy increment and the trapezoidal integral of the density-shape factor on every neighboring attractive interval. Midpoints are geometric in $\epsilon$. Open symbols mark states, or intervals with an endpoint, above the stationarity criterion. No derivative smoothing or fitted correction is used.}
\label{fig:energy}
\end{figure*}

The near-boundary points in Fig.~\ref{fig:energy}(a) have nearly linear squared energy. Their slopes reflect a slowly varying density-shape response. Fits to the three closest accepted points give $G_{\rm int}/\beta=\fitRootNineteen,\fitRootThirtySeven,\fitRootSixtyOne$, respectively, close to the known value $-4\pi=-12.56637$. Their role is a consistency check and a local description of the response. Full fit-window comparisons are retained in the SM.

More directly, Eq.~\eqref{eq:E2slope} predicts, along a smooth stationary branch,
\begin{equation}
 \Y(\kappa_2)-\Y(\kappa_1)
 =\int_{\kappa_1}^{\kappa_2}\bsh(\kappa)\,\dd\kappa.
 \label{eq:integralresponse}
\end{equation}
This compares energy increments with independently evaluated density moments and avoids differentiating sparsely sampled energies. We evaluate the integral by the trapezoidal rule on each neighboring interval and plot
\begin{equation}
 \mathcal R_i=\frac{\Delta\Y_i-\tfrac12(b_i+b_{i+1})\Delta\kappa_i}
 {\Delta\Y_i}.
 \label{eq:responseerror}
\end{equation}
Here $\Delta\Y_i=\Y(\kappa_{i+1})-\Y(\kappa_i)$, $\Delta\kappa_i=\kappa_{i+1}-\kappa_i$, and $b_i=\bsh(\kappa_i)$. The finite-interval discrepancy includes both branch nonstationarity and quadrature error from shape curvature; it is not a direct estimate of continuum error.

Over intervals whose endpoints both satisfy Eq.~\eqref{eq:criterion}, the largest absolute mismatches are $\responseMaxNineteen\%$, $\responseMaxThirtySeven\%$, and $\responseMaxSixtyOne\%$ for $\beta=19,37,61$. The $\beta=61$ value occurs between $\kappa=-0.236$ and zero, where the zero-coupling endpoint is the re-converged field rather than the continuation start. The only larger value, $2.8\%$ at $\beta=37$ between $\kappa=-0.969$ and $-0.945$, involves an endpoint above the criterion. Thus the contact-response identity is satisfied at the sub-percent level on accepted intervals, and the check independently flags the states that fail the stationarity criterion. Small violations of exact ground-state squared-energy concavity remain documented in the SM.

The physical interpretation is consequently based on the large change in magnetic defect, the accompanying density-shape evolution, and the spatial measurements below. A fitted zero is not needed to establish those observations.

\section{Spatial redistribution and the amplitude--flow balance}\label{sec:spatial}
\subsection{A positive local measure of magnetic nonsaturation}
We reconstruct $d(\rvec)=|(D_x-\mathrm{i}D_y)u|^2$ from all 14 stored attractive fields at $\beta=61$. For representative maps we use $\kappa=0,-0.709,-0.975$. In scaled coordinates $\boldsymbol{\xi}=\rvec/\sqrt P$, the normalized quantity
\begin{equation}
 \widehat d(\boldsymbol{\xi})=\frac{P}{Q}d(\sqrt P\,\boldsymbol{\xi}),
 \qquad \int_{\mathbb R^2}\widehat d(\boldsymbol{\xi})\,\dd\boldsymbol{\xi}=1
 \label{eq:normalizeddefect}
\end{equation}
separates the location of the remaining defect from its rapidly decreasing integrated magnitude. All integrals use the original grid; interpolation is confined to displaying the maps.

\begin{figure*}[t]
\centering
\includegraphics[width=\textwidth]{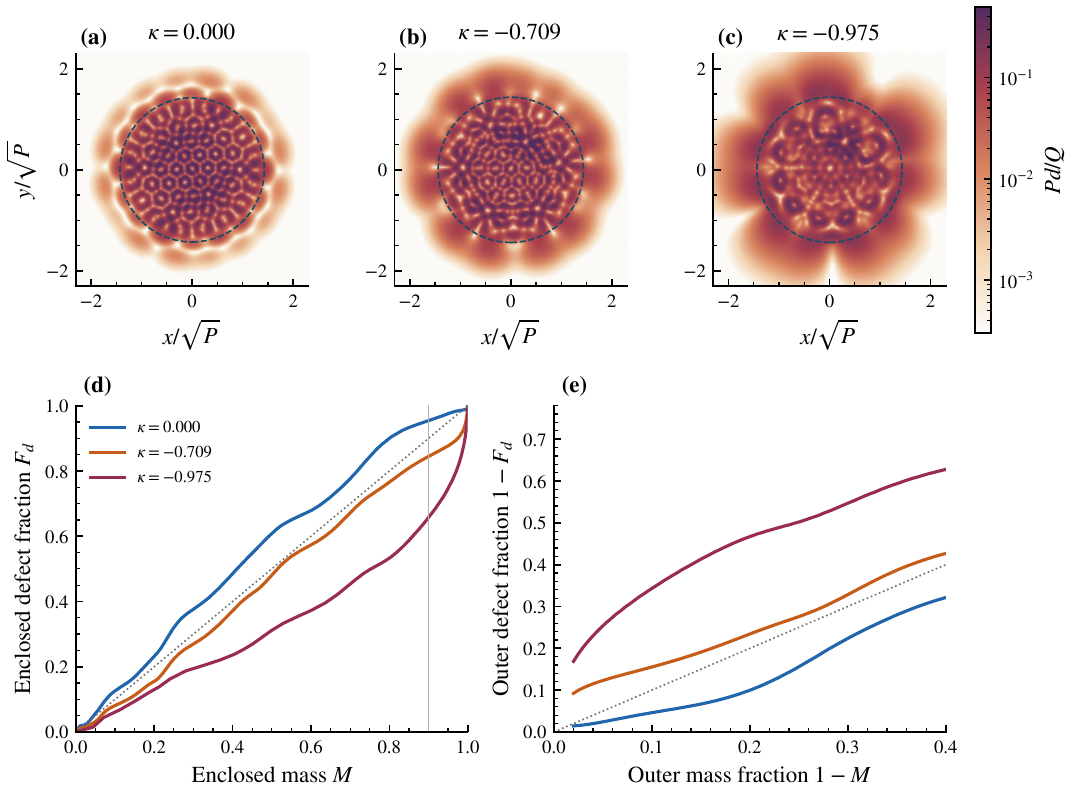}
\caption{Spatial organization of the self-duality defect at $\beta=61$. (a)--(c) Probability-normalized defect density $Pd/Q$ at zero coupling, moderate attraction, and the closest attractive state satisfying the stationarity criterion, with a shared logarithmic color scale. Dashed circles enclose $90\%$ of particle mass. (d) Enclosed defect versus enclosed mass. (e) Defect outside a variable mass radius versus the corresponding outer mass fraction. Dotted diagonals indicate equal radial distributions of mass and defect. The relative outward bias becomes pronounced under strong attraction.}
\label{fig:spatial}
\end{figure*}

The defect has structure on the vortex scale and a broader contribution towards the outer cloud. The cumulative curves quantify their combined radial weighting. At $\kappa=-0.975$, $F_d$ lies well below the mass diagonal over a broad interval. Outside $R_{0.9}$, the defect fraction changes from $0.046$ at $G=0$ to $0.155$ at $\kappa=-0.709$ and $0.344$ at $\kappa=-0.975$. The absolute defect nevertheless falls strongly: $Q$ decreases by a factor of about $13.7$, while $I$ and $T$ grow by factors of about $6.7$ and $6.5$.

\begin{table*}[t]
\caption{Representative reconstructed states at $\beta=61$. $Q_d$ is integrated directly from the local covariant defect. The outer fraction refers to the mass radius $R_{0.9}$. Digits specify the computed fields; the interpretation uses the size and evolution of the effect.}\label{tab:spatial}
\begin{ruledtabular}
\begin{tabular}{rrrrrrr}
$\kappa$ & $\sqrt P$ & $T$ & $I$ & $Q_d$ & $\delta$ & $1-F_d(R_{0.9})$\\
0 & 2.82849 & 8.00036 & 0.0201704 & 0.269584 & 0.0348715 & 0.04603\\
$-0.70898$ & 2.09635 & 14.86651 & 0.0385370 & 0.0962668 & 0.0065176 & 0.15525\\
$-0.97485$ & 1.14814 & 51.64986 & 0.1347081 & 0.0197042 & 0.0003816 & 0.34371\\
\end{tabular}
\end{ruledtabular}
\end{table*}

\subsection{Interaction dependence and independent spatial baselines}
\begin{figure*}[t]
\centering
\includegraphics[width=\textwidth]{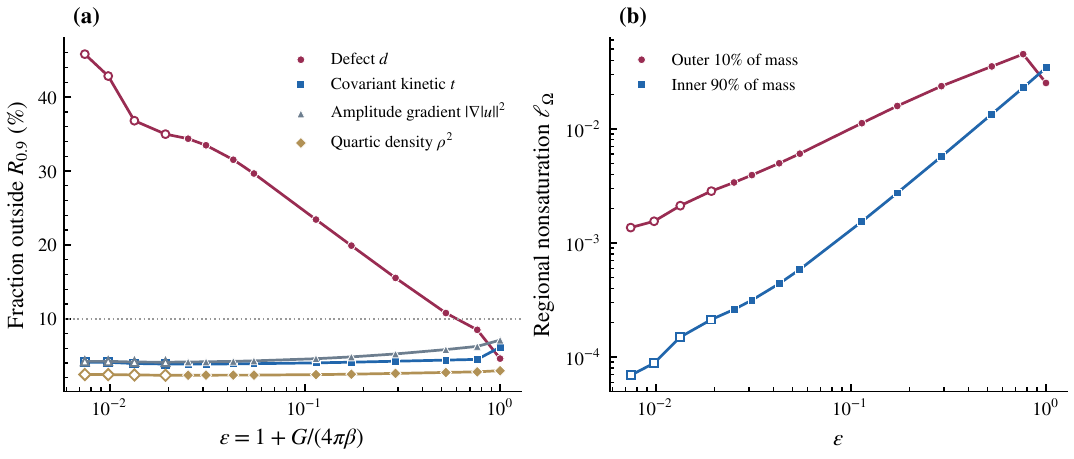}
\caption{Peripheral enrichment across all 14 attractive $\beta=61$ fields. (a) Fractions of the defect, covariant kinetic energy, amplitude-gradient energy, and quartic density outside the radius containing $90\%$ of mass. The horizontal reference marks $10\%$. (b) Regional nonsaturation $\ell_\Omega=(\int_\Omega d(\rvec)\,\dd\rvec)/(\int_\Omega t(\rvec)\,\dd\rvec)$ for the inner and outer regions. Both regions approach magnetic saturation under strong attraction, but the interior does so more efficiently. The regions are defined by mass, without assuming a sharp cloud support. Open symbols mark states above the stationarity criterion, Eq.~\eqref{eq:criterion}.}
\label{fig:regional}
\end{figure*}

Figure~\ref{fig:regional}(a) extends the outer-fraction measurement to the entire attractive sequence. The outer defect fraction grows monotonically along the whole attractive sequence: from $4.6\%$ at zero coupling, below the $10\%$ mass share, through $8.5\%$, $15.5\%$, $23.4\%$, and $29.7\%$ at $\kappa=-0.236,-0.709,-0.886,-0.945$, to $\outerDefectPercent\%$ at $\kappa=-0.975$. The four closer states, above criterion~\eqref{eq:criterion}, continue the rise to $\nearestDefectPercent\%$ (open symbols). At zero coupling the periphery is thus slightly less defective than its mass share, and the enrichment is produced by attraction.

The comparison with positive energy densities is informative. Define
\begin{equation}
 t(\rvec)=|D_xu|^2+|D_yu|^2,
 \qquad q(\rvec)=|\nabla |u||^2.
 \label{eq:localbaselines}
\end{equation}
At $\kappa=-0.975$, the outer tenth of particle mass carries $\outerDefectPercent\%$ of $d$, but only $\outerKineticPercent\%$ of $t$ and $\outerAmplitudePercent\%$ of $q$. Its fraction of $\rho^2$ is about $2.4\%$. The large defect fraction is consequently not explained by those positive energy densities simply accumulating in the same outer region. In fact, the outer amplitude-gradient fraction decreases from about $7.1\%$ at zero coupling to $4.2\%$ at $\kappa=-0.975$, while the outer defect fraction grows. The kinetic and amplitude-gradient fractions are essentially the same in the four less converged closest states, so the comparison does not rely on them.

The conclusion persists when the mass radius is varied. At $\kappa=-0.975$, the regions outside the $80\%$, $85\%$, and $95\%$ radii carry defect fractions about $0.467$, $0.412$, and $0.254$, compared with $0.100$, $0.067$, and $0.024$ at zero coupling. The measurement identifies relative peripheral enrichment, with a substantial defect contribution remaining inside each reference radius.

\subsection{Local cancellation of amplitude gradients and statistical flow}
To interpret the baseline comparison, write $u=s e^{\mathrm{i}\phi}$ away from its zeros, with $s=\sqrt\rho$, and define
\begin{equation}
 \bm v=\nabla\phi+\beta\vA,\qquad
 \bm J=\operatorname{Re}(\bar u\bm D u)=\rho\bm v.
 \label{eq:velocity}
\end{equation}
The current and $\bm v$ are gauge invariant on their domains. The exact local identities are
\begin{align}
 t&=|\nabla s|^2+s^2|\bm v|^2,\label{eq:localT}\\
 d&=|s\bm v+\nabla^\perp s|^2\nonumber\\
  &=|\nabla s|^2+s^2|\bm v|^2
       +2s\bm v\cdot\nabla^\perp s.
 \label{eq:localD}
\end{align}
Self-duality requires $\bm v=-\nabla^\perp\log s$. It is a relation between flow and density gradients: both positive contributions to $t$ can be large while their signed cross term nearly cancels them in $d$. The amplitude-gradient energy is therefore a useful comparison density, but is not a positive additive component of the defect by itself.

We measure the residual of this cancellation in a region $\Omega\subset\mathbb R^2$ by
\begin{equation}
 \ell_\Omega=\frac{\int_\Omega d(\rvec)\,\dd\rvec}{\int_\Omega t(\rvec)\,\dd\rvec}.
 \label{eq:regionalleakage}
\end{equation}
Globally, $Q/T=\delta/(1+\delta)$. Regionally, Fig.~\ref{fig:regional}(b) resolves different degrees of saturation. At $\kappa=-0.975$,
\begin{equation}
 \ell_{\rm in}\simeq2.6\times10^{-4},\qquad
 \ell_{\rm out}\simeq3.4\times10^{-3}.
 \label{eq:leakagenumbers}
\end{equation}
Thus the relative nonsaturation is about $\regionalLeakageRatio$ times larger in the outer region. At zero coupling the ratio is $0.7$, i.e., the interior is the less saturated region. The contrast is already a factor of about $10.3$ at $\kappa=-0.945$ and reaches about $\nearestLeakageRatio$ in the closest, less converged state. Both regional values decrease under strong attraction; the relative peripheral signal arises because cancellation is more effective in the interior. This provides a kinematic mechanism for the observed distribution without assuming that density gradients themselves are largest at the outer mass radius. It does not identify a unique dynamical process of vortex rearrangement.

The reconstruction checks Eqs.~\eqref{eq:localT} and \eqref{eq:localD} independently of the radial integrals. Amplitude/phase variables are undefined at exact zeros; the reported decomposition excludes such pixels and monitors their kinetic weight, below $3.1\times10^{-12}$ in these fields. The defect itself is always evaluated directly from $\bm D u$. Agreement of its integral with $T-2\pi\beta I$ and the available optimization diagnostics are reported in the SM.

\subsection{Radial envelope and angular density variation}
\begin{figure*}[t]
\centering
\includegraphics[width=\textwidth]{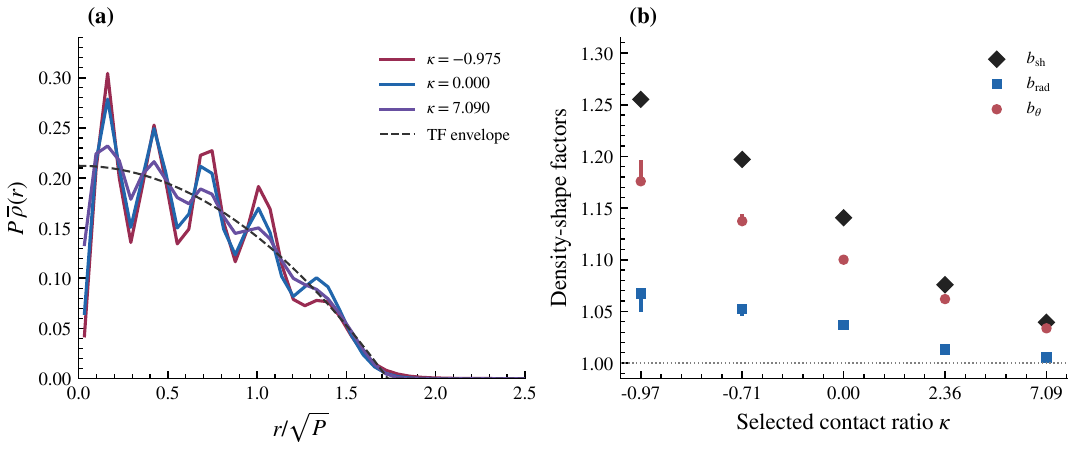}
\caption{Radial and angular density structure at $\beta=61$. (a) Scaled annular density profiles for strong attraction, zero coupling, and repulsion, with the fixed-second-moment Thomas--Fermi reference. Profile annuli have width $0.065\sqrt P$. (b) Decomposition $\bsh=\brad\btheta$. Couplings are equally spaced categories; markers use width $h$, and ranges span widths $h,2h,4h$. The ranges show annular discretization sensitivity. Angular density variation contributes substantially to the increased quartic cost under attraction.}
\label{fig:shape}
\end{figure*}

The shape response has a complementary spatial interpretation. Figure~\ref{fig:shape}(a) compares the scaled radial average with the fixed-second-moment Thomas--Fermi envelope,
\begin{equation}
 P\rho_{\rm TF}(r)=\frac{2}{3\pi}
 \left(1-\frac{r^2}{3P}\right)_+.
 \label{eq:scaledTF}
\end{equation}
The retained oscillations reflect vortex shells. Repulsion brings the broad profile closer to the reference, whereas attraction increases its modulation.

For width $h$, the zero-coupling factors are $\brad\simeq1.037$ and $\btheta\simeq1.100$, with product $1.141$. At $\kappa=-0.975$ they become about $1.067$ and $1.176$, with product $1.255$. Width $4h$ gives $1.051$ and $1.194$ at that point, preserving the product. Both radial structure and angular variation contribute, with the angular factor providing the larger enhancement in this decomposition.

Annular averaging retains radial shell oscillations and includes finite-envelope distortions in the angular variance. The decomposition therefore resolves rotational structure about the trap center rather than assigning a homogeneous vortex-cell energy. At the strongest repulsion, the factors approach $1.006$ and $1.034$, with weaker bin-width sensitivity. This smoother density coexists with a larger magnetic defect, as discussed next.

\subsection{Flux dependence, robustness, and the vortex-free periphery}\label{sec:robust}
A separate confirmation run with the same solver tests the spatial result in three ways (SM, Sec.~S7). First, the $\beta=19$ and $\beta=37$ attractive branches were recomputed with all fields stored. The $\beta=19$ branch reproduces the energies of Sec.~\ref{sec:attraction} and satisfies $r_{\rm rel}\leq2.4\times10^{-5}$ at every coupling. Its outer defect fraction rises from $7.7\%$ at zero coupling to $75.6\%$ at $\kappa=-0.975$, and $\ell_{\rm out}/\ell_{\rm in}$ from $0.6$ to $28$, while the outer kinetic fraction stays near $10\%$. The recomputed $\beta=37$ continuation followed a different, higher-energy branch than the one used in Sec.~\ref{sec:attraction} (by $0.15\%$ at zero coupling and up to $3\%$ near the boundary); on it the outer defect fraction rises from $12\%$ to $25\%$ and the ratio from $2.3$ to $10$. The enrichment is therefore present at all three fluxes, but its magnitude depends on flux and on the selected low-energy branch.

Second, at $\beta=61$ and $\kappa=-0.975$ the outer fraction is stable to further optimization ($34.4\%\to34.3\%$ over $10^4$ additional iterations), to a $1.25$ times finer grid ($34.2\%$), and to a $1.3$ times larger box ($34.2\%$), with relative energy changes below $6\times10^{-6}$. Two independent lattice starts placed directly at this coupling did not reach the criterion and ended $0.2\%$ and $1.3\%$ higher in energy, with outer fractions $46\%$ and $33\%$ and ratios $18$ and $11$; a random start remained far from stationarity. Third, the phase winding $W(R)=(2\pi)^{-1}\oint_{|\rvec|=R}\dd\phi$ on circles enclosing mass $M(R)$ locates the vortices; it is negative, i.e., the vortices wind opposite to the statistical flux. In every state $|W|$ saturates by $M\simeq0.8$: at $\beta=61$, $|W|=53$ at $\kappa=-0.975$ and $51$ at zero coupling, while the enclosed statistical flux keeps growing to $\beta M$. The interior is slightly over-screened, and the outer fifth of the mass is vortex free and carries $\beta-|W|\approx4$--$10$ unscreened flux units.

This vortex-free periphery suggests a kinematic origin for the enrichment. In polar coordinates Eq.~\eqref{eq:localD} splits exactly as $d=(sv_r-r^{-1}\partial_\theta s)^2+(sv_\theta+\partial_rs)^2$. For an approximately axisymmetric annulus without vortices, the circulation gives the azimuthal flow $v_\theta\simeq[\beta M(r)-|W|]/r$, and only the second term survives, $d\simeq s^2(v_\theta+\partial_r\log s)^2$. Local self-duality would then require $-r\partial_r\log s=\beta M(r)-|W|$, i.e., a power-law tail $s\propto r^{-(\beta-|W|)}$ once $M\simeq1$, as realized by the exact radial NLL of Appendix~\ref{app:NLL}, for which $\beta-|W|=n+1$. The stored fields test this directly (SM, Fig.~\ref{supp:fig:edge}). At $\kappa=-0.975$ the angular-averaged flow $r\bar v_\theta$ rises through the vortex-free shell from about $-1$ at $M=0.8$ to about $7.6$, close to $\beta-|W|=8$, and the logarithmic decay rate $-r\partial_r\log\bar s$ follows it to within $0.8$ out to $M=0.995$, separating only in the far tail where the trap dominates. The term $(sv_\theta+\partial_rs)^2$ carries $93\%$ of the defect outside $R_{0.8}$, and the axisymmetric profiles alone reproduce that outer defect to within $6\%$ ($78\%$ and a ratio of $0.78$ in the closest, unaccepted state; SM, Sec.~S8). At zero coupling the corresponding values are $57\%$ and $33\%$: there the outer defect is mostly nonaxisymmetric lattice structure. Strong attraction thus turns the peripheral defect into a smooth radial mismatch between statistical flow and density decay in a vortex-free shell, which the interior lattice cannot remove. As attraction saturates the interior, this contribution becomes the dominant part of the remaining defect. The analysis identifies the responsible term; it does not predict the size of the mismatch.

\section{Repulsion and the approach to a smoother density}\label{sec:repulsion}
\subsection{Magnetic defect and density shape move differently}
\begin{figure*}[t]
\centering
\includegraphics[width=\textwidth]{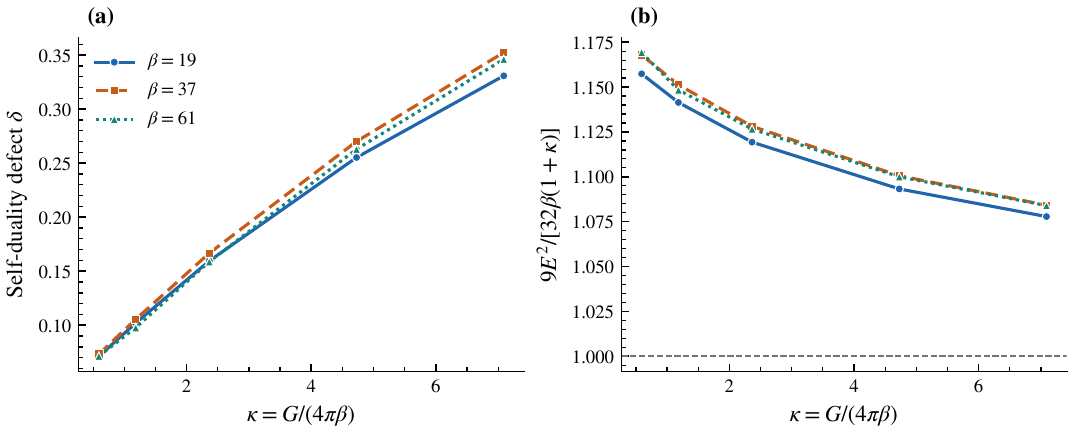}
\caption{Complementary repulsive response. (a) The relative magnetic defect increases with positive contact coupling. (b) The squared-energy ratio $\Y/\epsilon=9E^2/[32\beta(1+\kappa)]$ decreases towards its lower-bound reference. The ratio combines the density-shape and self-duality contributions, $\bsh(1+\eta)\mathcal V(w)$; it is not a direct homogeneous equation of state. The positive-coupling branches are selected independently of the attractive continuation, as specified in Sec.~\ref{sec:method}.}
\label{fig:repulsion}
\end{figure*}

Repulsion expands the cloud and suppresses the relative density depletion away from the immediate cores. It does not favor magnetic saturation in the same way as attraction. Figure~\ref{fig:repulsion}(a) shows that $\delta$ increases with $\kappa$ along all three positive-coupling branches. At the strongest sampled coupling, $\kappa\simeq7.090$, the values are $0.331$, $0.352$, and $0.346$ for $\beta=19,37,61$.

At the same couplings, $\bsh$ decreases to $1.0355$, $1.0389$, and $1.0395$. There is no contradiction: the interaction penalizes the full quartic density, whereas $\delta$ compares the kinetic energy with the magnetic quartic value of that changing density. A nearly smooth density can require greater excess covariant kinetic energy than a more strongly modulated, nearly self-dual field. The annular decomposition in Fig.~\ref{fig:shape} resolves the corresponding approach of the envelope and azimuthal variance factors towards one.

The measured energy response is summarized by
\begin{equation}
 \frac{\Y}{\epsilon}=\bsh(1+\eta)\mathcal V(w)\geq1.
 \label{eq:responseRatio}
\end{equation}
Figure~\ref{fig:repulsion}(b) shows a decrease with repulsion. At the strongest coupling, the ratio is approximately $1.078$--$1.084$ across the three fluxes. The residual enhancement above one therefore contains both the remaining density modulation and the positive magnetic defect. Replacing this combination by a constant cell factor would obscure their different evolution.

\subsection{Fixed-flux strong-repulsion limit}
At fixed finite $\beta$, the leading strong-repulsion limit can be understood directly from scaling. Put $u(\rvec)=G^{-1/4}\tilde u(\rvec/G^{1/4})$. Then
\begin{equation}
 \frac{\mathcal E_{\beta,G}[u]}{\sqrt G}
 =\frac{T_\beta[\tilde u]}{G}+\frac12 I[\tilde u]+P[\tilde u].
 \label{eq:repulsiveTFscale}
\end{equation}
Here $T_\beta$ is the kinetic energy at flux $\beta$, which scales as an inverse length squared.
The kinetic term is nonnegative, giving the contact--trap density functional as a lower bound. Smooth approximations to its minimizing density give the matching leading upper bound for fixed $\beta$. Thus
\begin{equation}
 E_{\rm gs}(\beta,G)=\frac43\sqrt{\frac{G}{2\pi}}\,[1+o(1)],
 \qquad G\to+\infty\quad(\beta\ \text{fixed}).
 \label{eq:strongrep}
\end{equation}
This elementary fixed-flux statement concerns the leading energy, not the optimal vortex arrangement or a joint large-$\beta$ limit. It predicts $\Y/\epsilon\to1$, consistently with the direction of Fig.~\ref{fig:repulsion}(b), without establishing that the last numerical coupling is asymptotic.

The triangular Abrikosov value $1.1596$ belongs to a distinct constrained cell problem in the appropriate lowest-Landau-level regime \cite{KleinerRothAutler1964,CooperKomineasRead2004,AftalionBlancDalibard2005}. A triangular arrangement of small cores does not make every global density factor equal to that number. In the present data, the radial and angular factors both decrease under strong repulsion. Establishing an interacting homogeneous coefficient would require a separate thermodynamic or magnetic-cell calculation with the correct nonzero mean flux; it cannot be inferred from the trapped energy ratio alone.

\section{Variational interpretation of the contact response}\label{sec:variational}
The dilation representation also explains the usefulness and limitations of nearly straight squared-energy curves. A stored shape $u_i$ supplies the scale-optimized trial energy
\begin{equation}
 U_i(G)^2=4P_i(T_i+GI_i/2).
 \label{eq:trialline}
\end{equation}
For a finite family, the best such trial response is
\begin{equation}
 U_{\mathcal F}(G)^2=\min_{i\in\mathcal F}U_i(G)^2,
 \label{eq:finiteenvelope}
\end{equation}
a concave piecewise-affine function. In dimensionless form, each shape gives
\begin{equation}
 \Y_i(\kappa)=b_i(1+\kappa+\delta_i),\qquad
 b_i=\frac{9\pi}{4}I_iP_i.
 \label{eq:trialY}
\end{equation}
Both $b_i$ and $\delta_i$ are invariant under dilation. A fixed nonsaturating shape therefore has its extrapolated zero at $\kappa=-1-\delta_i$. A slightly over-attractive fitted intercept can reflect a small positive shape defect, while the exact stability line remains unchanged.

Selecting the best stored shape gives a useful internal comparison without new minimization. At $\beta=37$, $\kappa=-0.945$, the shape from $\kappa=-0.969$ lowers the source energy by about $0.65\%$ after rescaling; this is the $\beta=37$ state flagged by criterion~\eqref{eq:criterion}, so the construction and the residual identify the same weak point independently. Elsewhere the improvement is at most $0.2\%$ ($\beta=37$, $\kappa=-0.886$, again using the $\kappa=-0.969$ shape). These modest improvements delimit the energy accuracy of the selected continuation states while leaving the much larger attractive change in the defect as a distinct observation. The full finite-family comparison is shown in the SM. With discrete energies it is a numerical variational construction, rather than a certified continuum bound.

Exact NLL trial shapes provide a complementary benchmark \cite{JackiwPi1990PRD,AtaeiLundholmNguyen2024,AtaeiEtAl2025}. The radial ring at $\beta=2n$ has $b_{\rm sh}\sim3n/8$, whereas the sampled lattice states have order-one shape factors. Appendix~\ref{app:NLL} gives the exact field and moments. This geometric comparison illustrates that magnetic saturation alone does not select a low-cost density shape; the even-flux rings and odd-flux numerical states are not matched-flux energy competitors.

\section{Discussion}\label{sec:discussion}
\subsection{Spatially nonuniform approach to self-duality}
The attractive lattice response has three distinct components. The cloud contracts, the magnetic defect decreases, and the full-density shape factor increases. Since $\delta$ and $\bsh$ are dilation invariant, their evolution reflects a reorganization beyond a change of cloud radius. The angular density variation can grow while the flow--amplitude balance approaches the self-dual condition. A modulated density and a small magnetic defect are therefore compatible properties of the same field.

The central spatial result is the increasing distinction between the interior and periphery under strong attraction. A small global defect hides a nonuniform degree of cancellation. In the closest accepted $\beta=61$ state, the outer tenth of mass carries about one third of the remaining defect, despite carrying only about one twenty-fifth of the covariant kinetic energy. Its relative nonsaturation is about thirteen times the interior value, and both numbers keep growing in the less converged states closer to the boundary. The positive amplitude-gradient baseline has no comparable peripheral enhancement. The regional observable therefore distinguishes the spatial failure of self-dual cancellation from the location of the underlying positive kinetic cost.

This is an energetic characterization of computed stationary-state approximations. It neither requires a sharp edge nor asserts complete saturation in the interior. The mass-radius construction remains meaningful when the density has tails and discrete vortex shells. With the re-converged zero-coupling reference, the outer defect fraction grows monotonically with attraction; the change is a crossover in the present measurements, rather than evidence for a phase transition.

\subsection{Contribution relative to prior anyon and NLL results}
The stability line and the even-flux exact NLL classification are established inputs \cite{AtaeiLundholmNguyen2024}. Trapped CSS spectra, contact interactions of both signs, and NLL variational comparisons are already treated in Ref.~\cite{AtaeiEtAl2025}. Our focus is the spatial organization of nonsaturation in the resolved vortex lattice: its contact dependence, its comparison with amplitude and kinetic energy densities, and the associated density-shape response.

The scale and contact identities supply the organizing framework. They explain why a squared-energy slope measures density geometry and why an accurate-looking fitted zero need not measure a new threshold. Their integral form gives a direct finite-interval check between energies and density moments. The amplitude--flow identity then connects the integrated language to the local maps. The contribution lies in using these relations together to resolve the changing distribution of a small defect.

This regime differs from weak-field attractive collapse near the Townes problem \cite{Nguyen2024}, from fixed-statistics ideal-anyon estimates \cite{LundholmSeiringer2018}, and from fermionic-gauge magnetic Thomas--Fermi theory \cite{LevittLundholmRougerie2026}. Those theories address different limits or effective descriptions. In particular, the trapped shape factor here combines envelope and vortex modulation and should be distinguished from the Abrikosov cell ratio of a prescribed-field lowest-Landau-level problem.

\subsection{Numerical scope and endpoint questions}
The appropriate numerical precision depends on the observable. The decrease in $\delta$ by about two orders of magnitude appears across three fluxes. The peripheral response is resolved over all 14 attractive $\beta=61$ fields, and every quoted value comes from states satisfying Eq.~\eqref{eq:criterion}. It is already substantial well inside the accepted range: at $\kappa=-0.945$ the outer defect fraction is about $30\%$ with $r_{\rm rel}=3.5\times10^{-5}$, and at $\kappa=-0.975$ it is $\outerDefectPercent\%$ with $r_{\rm rel}=4.1\times10^{-4}$. The contrast with the roughly $4\%$ outer kinetic fraction is much larger than the last digits of the energy or fitted root.

The exact endpoint and global-minimizer problem require stronger control. Five of the 36 attractive states exceed the practical criterion, and the tightest requested tolerance is reached only away from the boundary. The confirmation run shows that the regional fractions are converged with respect to grid, box, and further optimization at the $0.2$ percentage-point level, but that other low-energy branches and starts change them by up to about ten percentage points. We therefore report the regional fractions as properties of the lowest-energy branches found, rather than as universal constants; the robust statements are the enrichment itself and the large ratio $\ell_{\rm out}/\ell_{\rm in}$ near the boundary. The response mismatches, annular-width sensitivity, and local identity checks specify what has been tested; the SM records the remaining diagnostics without smoothing away their discrepancies.

The odd principal fluxes also matter for endpoint claims. Regular exact saturation occurs at even flux, while $19,37,61$ are odd. A matched even--odd sequence would address the relation between finite-flux concentration and exact NLL trial families. It is an extension of the present spatial study, rather than an assumption needed to define its observables. The current results do not determine a universal critical shape factor or the trapped energy exactly at the odd-flux boundary.

\subsection{Physical setting and extensions}
The CSS functional is an effective target model. Quantum Hall collision and braiding experiments probe fractional statistics in electronic quasiparticles \cite{Bartolomei2020,Nakamura2020}. Density-dependent gauge couplings and topological gauge dynamics have also been realized in optically dressed Bose gases \cite{YaoZhangChin2022,Frolian2022}, while attractive two-dimensional gases exhibit vortex-collapse dynamics \cite{BanerjeeEtAl2025}. These developments motivate studying statistical feedback and attraction together, although an isotropic two-dimensional implementation of the present functional remains a separate experimental task.

The regional diagnostics can be applied to nonradial NLL trial families, different trapping profiles, and improved even--odd flux sequences. A homogeneous interacting equation of state is another useful extension, with nonzero mean flux and gauge-periodic boundary conditions retained explicitly. The present trapped analysis provides spatial and response constraints for those calculations.

\section{Conclusion}\label{sec:conclusion}
Contact attraction drives the computed anyon vortex lattices towards magnetic saturation while increasing density modulation. The relative defect decreases from $3.5$--$4\%$ to $1.3\times10^{-4}$--$3.8\times10^{-4}$ at $\beta=19,37,61$ in states satisfying our stationarity criterion (at $\epsilon=0.0074$ for $\beta=19,37$ and $\epsilon=0.025$ for $\beta=61$). Reconstruction across the full attractive $\beta=61$ sequence reveals a spatially nonuniform approach: under strong attraction, the residual defect becomes progressively enriched in the outer cloud.

Comparing the defect with covariant kinetic and amplitude-gradient energy densities resolves the meaning of that enrichment. In the closest accepted state at $\beta=61$, the outer tenth of mass carries about $34\%$ of the defect but only about $4\%$ of either positive kinetic baseline. Its relative nonsaturation is about thirteen times that of the interior; at $\beta=19$ the corresponding values are $76\%$ and $28$. The outer fifth of the mass is vortex free and carries a few unscreened flux units; under strong attraction the defect there is almost entirely a small radial mismatch between the statistical flow and the density decay, which the interior lattice cannot cancel. The magnitude of the enrichment depends on flux and on the selected low-energy branch, whereas its presence and the large regional contrast are robust. Annular density moments resolve the accompanying growth of angular modulation. Repulsion gives the complementary response of smoother density and a larger magnetic defect.

Dilation and contact-response identities connect the spatial picture with squared energy and provide finite-interval consistency checks. Together, these observables characterize how a modulated many-vortex field approaches self-duality and identify where its remaining departure is concentrated, with numerical scope specified by the accompanying diagnostics.

\section*{ACKNOWLEDGMENTS}
We sincerely thank Dr. Dinh-Thi Nguyen for his invaluable academic insights and, above all, for being a constant source of inspiration throughout the writing of this article. 

\section*{Data availability}
The complex fields, solver, confirmation and post-processing scripts, and figure data are provided in a reproducibility archive upon request.  

\appendix
\section{Density inequality and virial correction}\label{app:density}
For unit mass at fixed $P$, minimization of $I=\int_{\mathbb R^2}\rho(\rvec)^2\,\dd\rvec$ over nonnegative densities gives
\begin{equation}
 \rho_*(r)=\frac{2}{\pi R^2}\left(1-\frac{r^2}{R^2}\right)_+,
 \quad P=\frac{R^2}{3},\quad I_* =\frac{4}{3\pi R^2}.
 \label{eq:TFdensity}
\end{equation}
The Euler equation is affine in $r^2$ on the support. Convexity of $I$ gives the minimum, hence $IP\geq4/(9\pi)$. This is a density inequality; the equality of densities does not assert finite kinetic energy of $\sqrt{\rho_*}$ at its sharp boundary. Smooth approximations are sufficient for the variational limiting arguments.

Let $\mathcal T_G=Q+aI=(1+w)P$. Then $E=\mathcal T_G+P=(2+w)P$ and
\begin{equation}
 \frac{E^2}{4\mathcal T_GP}=\frac{(2+w)^2}{4(1+w)}
 =1+\frac{w^2}{4(1+w)}.
\end{equation}
Substituting $\mathcal T_G=2\pi\beta(\epsilon+\delta)I$ and $IP=4\bsh/(9\pi)$ gives Eq.~\eqref{eq:shapeidentity}. This calculation also shows that optimizing only the scale lowers a source energy by the factor $\mathcal V(w)^{-1/2}$. A small virial residual therefore controls that one improvement direction quadratically, not all possible deformations of the field.

For a differentiable stationary branch $u(G)$, the derivative of the constrained energy is its explicit $G$ derivative: terms involving $\dd u/\dd G$ vanish by the stationary equation and mass conservation. This gives Eq.~\eqref{eq:HF}; combining it with virial gives Eq.~\eqref{eq:E2slope}. The argument does not require global minimization, whereas the concavity constraint does.

\section{Radial NLL moments}\label{app:NLL}
An exact normalized radial NLL at $\beta=2n$ is
\begin{equation}
 v_n(r,\theta)=\sqrt{\frac n\pi}\frac{r^{n-1}e^{-\mathrm{i}(n-1)\theta}}{1+r^{2n}},
 \quad n\geq1.
 \label{eq:radialNLL}
\end{equation}
For $n>1$, its moments and shape factor are
\begin{align}
 P_n&=\frac{\pi}{n\sin(\pi/n)},\qquad
 I_n=\frac{1-n^{-2}}{6\sin(\pi/n)},\label{eq:NLLmoments}\\
 b_n^{\rm ring}&=\frac{3\pi^2(n^2-1)}{8n^3\sin^2(\pi/n)}
 =\frac{3n}{8}+O(n^{-1}).\label{eq:ringfactor}
\end{align}
The density of this field is
\begin{equation}
 \rho_n(r)=\frac n\pi\frac{r^{2n-2}}{(1+r^{2n})^2}.
\end{equation}
With $\tau=r^{2n}$, its mass is $\int_0^\infty(1+\tau)^{-2}\,\dd\tau=1$, and its enclosed mass is $M(r)=r^{2n}/(1+r^{2n})$. The second moment and quartic integral become beta-function integrals,
\begin{align}
 P_n&=B(1+n^{-1},1-n^{-1}),\\
 I_n&=\frac n\pi B(2-n^{-1},2+n^{-1}).
\end{align}
Using $\Gamma(1+z)\Gamma(1-z)=\pi z/\sin(\pi z)$ and $\Gamma(4)=6$ gives Eq.~\eqref{eq:NLLmoments}. The condition $n>1$ is required for $P_n$: the $n=1$ density decays as $r^{-4}$ and has a logarithmically divergent second moment. The large-$n$ expansion of the sine gives Eq.~\eqref{eq:ringfactor}. These moments provide an exact test for numerical radial quadrature and show explicitly which self-dual trial states can be inserted directly into a harmonic-trap upper bound.

For general rational NLL data, the leading tail depends on the degree of the Wronskian relative to the denominator. Finite mass does not by itself guarantee finite $P$. A trapped variational construction must either restrict to finite-moment states or control a tail modification together with its kinetic defect. Treating a divergent moment as a finite-grid number would spuriously improve the apparent trial bound.

\section{Free-space convolution and derivative checks}\label{app:kernel}
On the physical square of side $L$, all source--target separations are shorter than $\sqrt2L$. The logarithmic kernel is truncated at $R_{\rm c}=1.0001\sqrt2L$ and applied on a padded square of side $3L$. Its Fourier transform is
\begin{equation}
 \widehat\Phi(k)=-2\pi\left[\frac{1-J_0(R_{\rm c}k)}{k^2}
 -\frac{R_{\rm c}\log R_{\rm c}\,J_1(R_{\rm c}k)}{k}\right],
 \label{eq:truncatedkernel}
\end{equation}
where $J_0,J_1$ are Bessel functions and $k=|\bm k|$, with $\widehat\Phi(0)=-\pi R_{\rm c}^2(1-2\log R_{\rm c})/2$. The vector-potential multipliers are $-\mathrm{i}k_y\widehat\Phi$ and $\mathrm{i}k_x\widehat\Phi$. The density is zero padded, convolved, and restricted to the original square. Nyquist rows and columns are omitted consistently to preserve a real antisymmetric kernel. This is the truncated-kernel free-space strategy of Ref.~\cite{VicoGreengardFerrando2016}; no mean density is subtracted.

For a normalized Gaussian of width $\sigma>0$, $\rho=e^{-r^2/\sigma^2}/(\pi\sigma^2)$, the exact potential is
\begin{equation}
 \vA[\rho](\rvec)=\frac{1-e^{-r^2/\sigma^2}}{r^2}(-y,x).
\end{equation}
This checks the kernel, sign, and normalization independently of a many-vortex solution. A finite-difference directional derivative checks Eq.~\eqref{eq:gradient}. To see the origin of its backreaction sign, vary the field-dependent potential in $T$:
\begin{equation}
 \delta T\big|_{\vA}=2\beta\int_{\mathbb R^2} \bm J(\rvec)\cdot
 [\bm K*\delta\rho](\rvec)\,\dd\rvec.
\end{equation}
Oddness of $\bm K$ moves the convolution to $-\bm K*\bm J$ in the density variation. The coefficient of $\delta\bar u$ is therefore the final term of Eq.~\eqref{eq:gradient}.

\section{Discrete annular identity and reproducibility}\label{app:annuli}
Let an annulus $\mathcal C_j$ contain $n_j$ grid points, and let $S_j=\sum_{\rvec\in \mathcal C_j}\rho_{\rvec}$. The discrete radial mean is $\bar\rho_j=S_j/n_j$. With cell area $h^2$,
\begin{equation}
 I_{{\rm rad},h}=h^2\sum_j\frac{S_j^2}{n_j},\qquad
 I_{{\rm var},h}=h^2\sum_j\sum_{\rvec\in \mathcal C_j}(\rho_{\rvec}-\bar\rho_j)^2.
\end{equation}
The exact discrete identity is $I_h=I_{{\rm rad},h}+I_{{\rm var},h}$. Multiplying and dividing by $I_{{\rm rad},h}$ gives the product in Eq.~\eqref{eq:radialfactors} with the original discrete $P$. Finite annulus width does not preserve every pointwise radial moment exactly, so the continuum lower bound on $\brad$ is interpreted with that discretization in mind. The global product and nonnegative variance are exact for the arrays.

The reconstruction archive includes fixed checkpoints, source hashes, complex fields, and scripts for all figures. The SM and accompanying README specify the implementation checks and execution order.

\bibliography{references}

\clearpage
\onecolumngrid
\setcounter{equation}{0}
\setcounter{figure}{0}
\setcounter{table}{0}
\renewcommand{\theequation}{S\arabic{equation}}
\renewcommand{\thefigure}{S\arabic{figure}}
\renewcommand{\thetable}{S\arabic{table}}
\renewcommand{\theHequation}{supp.\arabic{equation}}
\renewcommand{\theHfigure}{supp.\arabic{figure}}
\renewcommand{\theHtable}{supp.\arabic{table}}
\setcounter{page}{1}
\renewcommand{\thepage}{S\arabic{page}}
\begin{center}
{\Large\bfseries Supplemental Material}\par\medskip
{\large Near self-duality and spatial reorganization of anyon vortex lattices}\par\smallskip
Bao-Duy Le\par\smallskip
{\small\itshape Faculty of Physics and Engineering Physics, University of Science, Ho Chi Minh City, Vietnam\\
Vietnam National University, Ho Chi Minh City, Vietnam}
\end{center}
 
\section*{S1. Definitions and coupling conventions}
For a normalized field on $\mathbb R^2$, with $\rvec=(x,y)$ and $r=|\rvec|$, we write
\begin{equation}
 \begin{gathered}
 E=T+\frac G2 I+P,\\
 T=\int_{\mathbb R^2}|(-\mathrm{i}\nabla+\beta\vA[|u|^2])u|^2\,\dd\rvec,\quad
 I=\int_{\mathbb R^2}|u|^4\,\dd\rvec,\quad
 P=\int_{\mathbb R^2}|\rvec|^2|u|^2\,\dd\rvec.
 \end{gathered}
\end{equation}
Here $\dd\rvec$ is two-dimensional Lebesgue measure, $\vA[\rho]=\nabla^\perp(\log|\cdot|*\rho)$, and $\nabla^\perp=(-\partial_y,\partial_x)$. The local defect is $d=|(D_x-\mathrm{i}D_y)u|^2$, its integral is $Q=\int_{\mathbb R^2}d(\rvec)\,\dd\rvec$, and $\delta=Q/(2\pi\beta I)$. The full-density shape factor is $b_{\rm sh}=9\pi IP/4$.

The numerical coupling label is $\gamma=G/(2e_{\rm Ab}\beta)$, with $e_{\rm Ab}\simeq7.42444$. Its relation to the variables used in the main text is $\kappa=e_{\rm Ab}\gamma/(2\pi)$ and $\epsilon=1+\kappa$. The nearest sampled coupling, $\gamma=-0.84$, has $\epsilon\simeq0.00743$. The established stability boundary is $G_c=-4\pi\beta$ for $\beta\geq2$, while regular exact self-dual states occur at even flux \citep{AtaeiLundholmNguyen2024}. These are analytical inputs to the numerical study.

\section*{S2. Resolution and stationarity}
Each attractive flux sequence uses one fixed continuation branch. Table~\ref{supp:tab:branches} gives the numerical resolution; $N_{\rm g}$ is the number of grid points per direction and $L$ is the square side in trap units. The free-space convolution and its padding are described in the main text.
\begin{table}[!htbp]
\centering
\caption{Resolution and number of attractive states; parentheses give the number satisfying $r_{\rm rel}\leq10^{-3}$.}\label{supp:tab:branches}
\begin{tabular}{rrrr}
\toprule
$\beta$ & $N_{\rm g}$ & $L$ & States\\
\midrule
19 & 384 & 13.819 & 11 (11)\\
37 & 512 & 15.597 & 11 (10)\\
61 & 640 & 17.205 & 14 (10)\\
\bottomrule
\end{tabular}
\end{table}

\begin{figure}[ht]
\centering
\includegraphics[width=.95\textwidth]{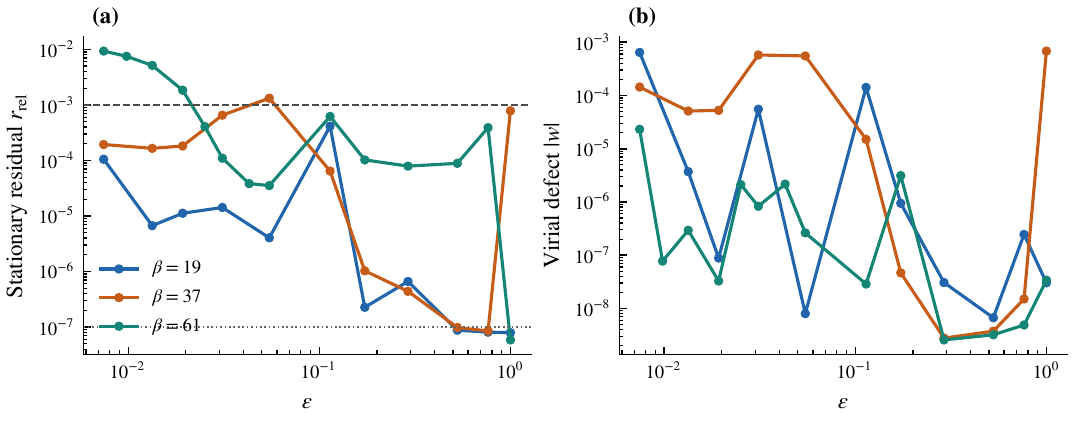}
\caption{Stationary and virial residuals along the attractive branches. In (a), the dashed line marks the practical criterion $10^{-3}$ and the dotted line the requested optimizer tolerance $10^{-7}$. The two residuals probe different variations of the field.}\label{supp:fig:residuals}
\end{figure}

Stationarity is measured by the relative residual $r_{\rm rel}$ defined in the main text. The optimizer requests $10^{-7}$; quantitative statements use the practical criterion $r_{\rm rel}\leq10^{-3}$. The virial deviation is $w=(T+GI/2-P)/P$. Figure~\ref{supp:fig:residuals} shows both diagnostics along the branches; representative values appear in Table~\ref{supp:tab:diagnostics}. Dilation relaxation can reduce $w$ without making the field stationary against other deformations. States above the practical criterion are marked by daggers and open symbols and are not used in fits or quoted values.

\begin{table}[ht]
\centering\small
\caption{Representative energy and stationarity diagnostics at the closest attractive coupling, the closest accepted coupling (where different), and zero coupling. A dagger marks $r_{\rm rel}>10^{-3}$. Displayed digits are not uncertainty estimates.}\label{supp:tab:diagnostics}
\begin{tabular}{rrrrrr}
\toprule
$\beta$ & $\kappa$ & $E$ & $\delta$ & $|w|$ & $r_{\rm rel}$\\
\midrule
\diagnosticrows
\bottomrule
\end{tabular}
\end{table}

The $\beta=61$ zero-coupling entry is the re-converged field of the confirmation run (Sec.~S7), which replaces the original continuation start. The branches are thus not certified global minimizers. Available box and grid comparisons also use unequally optimized states and do not provide a controlled continuum uncertainty.

\section*{S3. Contact response and fit-window sensitivity}
Along a differentiable minimizing branch, concavity and Hellmann--Feynman imply, for $G_1<G_2$,
\begin{equation}
 \frac{I(G_2)}2\leq\frac{E(G_2)-E(G_1)}{G_2-G_1}\leq\frac{I(G_1)}2.
\end{equation}
All adjacent intervals of the selected branches satisfy these secant inequalities. A small local violation of the stronger squared-energy concavity constraint remains at $\beta=37$, at the flagged state, consistent with incomplete optimization or local branch selection.

The integral response check compares $\Delta\mathcal Y$ with $(b_i+b_{i+1})\Delta\kappa/2$, where $\mathcal Y=9E^2/(32\beta)$ and $b_i=b_{\rm sh}(\kappa_i)$. Over intervals with both endpoints accepted, the largest relative mismatches are $\responseMaxNineteen\%$, $\responseMaxThirtySeven\%$, and $\responseMaxSixtyOne\%$ for $\beta=19,37,61$; the only interval with a flagged endpoint reaches $2.8\%$ ($\beta=37$). They include finite-interval quadrature and nonstationarity, and are not bounds on the continuum error or on a small defect obtained by subtraction.

Near the boundary, we fit $E^2=\sigma_{\rm fit}G+c_0$ and define $G_{\rm int}=-c_0/\sigma_{\rm fit}$ and $b_{\rm fit}=9\pi\sigma_{\rm fit}/8$. Only states satisfying $r_{\rm rel}\leq10^{-3}$ enter the fits. Table~\ref{supp:tab:fits} compares the closest two and three accepted points with all accepted points in the window $\gamma\leq-0.8$. The meshes differ between fluxes, so these fits are local consistency checks, rather than a matched-flux extrapolation or a measurement of a universal critical coefficient.
\begin{table}[ht]
\centering
\caption{Dependence of the squared-energy fit on the coupling window.}\label{supp:tab:fits}
\begin{tabular}{rrrrl}
\toprule
$\beta$ & Points & $G_{\rm int}/\beta$ & $b_{\rm fit}$ & $\gamma$ window\\
\midrule
19 & 2 & -12.57149 & 1.2862 & $[-0.840,-0.835]$\\
19 & 3 & -12.57191 & 1.2819 & $[-0.840,-0.830]$\\
19 & 5 & -12.57407 & 1.2669 & $[-0.840,-0.800]$\\
37 & 2 & -12.56832 & 1.2687 & $[-0.840,-0.835]$\\
37 & 3 & -12.56858 & 1.2659 & $[-0.840,-0.830]$\\
37 & 4 & -12.56894 & 1.2630 & $[-0.840,-0.820]$\\
61 & 2 & -12.57157 & 1.2536 & $[-0.825,-0.820]$\\
61 & 3 & -12.57245 & 1.2505 & $[-0.825,-0.810]$\\
61 & 4 & -12.57321 & 1.2480 & $[-0.825,-0.800]$\\
\bottomrule
\end{tabular}
\end{table}

\section*{S4. Spatial reconstruction and radial statistics}
All spatial observables are evaluated directly from the complex fields using the saturating covariant derivative $D_x-\mathrm{i}D_y$. For the three representative maps, the relative difference between $\int_{\mathbb R^2}d(\rvec)\,\dd\rvec$ and $T-2\pi\beta I$, normalized by $2\pi\beta I$, is below $4\times10^{-9}$. At $\kappa=-0.975$, the difference is about $3\times10^{-10}$ of $Q$ itself. This checks the reconstruction identity independently of the stationary residual.

Cumulative integrals of mass and defect define $M(R)$ and $F_d(R)$. The radius $R_\nu$ is the first grid radius enclosing mass fraction $\nu$; the outer region uses $r>R_\nu$. Integrals use the original grid, without display interpolation. At $\kappa=-0.975$, the defect fractions outside the radii enclosing $80\%$, $85\%$, $90\%$, and $95\%$ of mass are approximately $0.467$, $0.412$, $0.344$, and $0.254$; in the closest, unaccepted state at $\kappa=-0.993$ they are $0.564$, $0.515$, $0.458$, and $0.380$. At zero coupling they are $0.100$, $0.067$, $0.046$, and $0.024$. Peripheral enrichment therefore persists across several mass-radius choices.

\begin{figure}[ht]
\centering
\includegraphics[width=.95\textwidth]{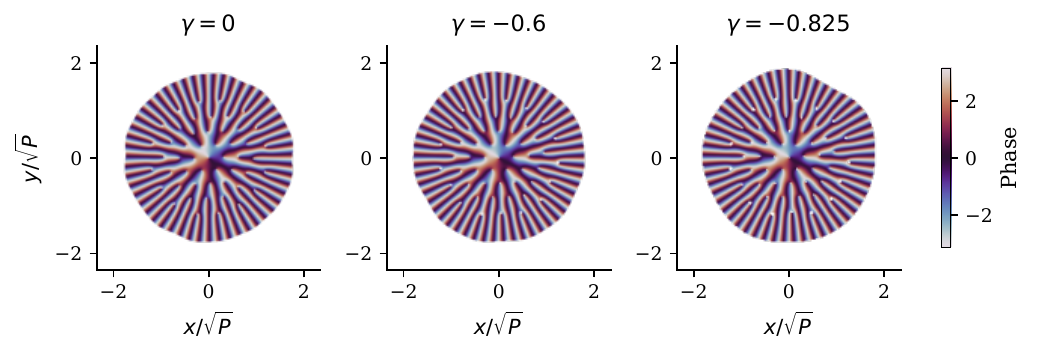}
\caption{Phase of the three representative states, masked where $\rho<0.01\rho_{\max}$. Coordinates are scaled by $\sqrt P$ as in the main text.}\label{supp:fig:phase}
\end{figure}

Annular averaging separates the radial mean from angular variance. Table~\ref{supp:tab:annuli} compares widths $h$ and $4h$, where $h=L/N_{\rm g}$. The exact discrete variance identity preserves $b_{\rm sh}=b_{\rm rad}b_\theta$, although the individual factors depend on bin width. The phase maps in Fig.~\ref{supp:fig:phase} provide a complementary view of the resolved vortex pattern.
\begin{table}[ht]
\centering
\caption{Sensitivity of the radial and angular factors to annular width at $\beta=61$. Bin-width variation is not a statistical error bar.}\label{supp:tab:annuli}
\begin{tabular}{rrrr}
\toprule
$\kappa$ & $b_{\rm sh}$ & $(b_{\rm rad},b_\theta)$, width $h$ & Width $4h$\\
\midrule
$-0.975$ & 1.255 & $(1.067,1.176)$ & $(1.051,1.194)$\\
$-0.709$ & 1.197 & $(1.053,1.137)$ & $(1.048,1.143)$\\
$0$ & 1.141 & $(1.037,1.100)$ & $(1.035,1.102)$\\
$2.363$ & 1.076 & $(1.013,1.062)$ & $(1.012,1.063)$\\
$7.090$ & 1.039 & $(1.006,1.034)$ & $(1.005,1.034)$\\
\bottomrule
\end{tabular}
\end{table}

\section*{S5. Amplitude--flow balance and the full spatial sequence}
With $s=|u|>0$, amplitude gradients and the gauge-invariant flow are reconstructed as
\begin{equation}
 \partial_j s=\frac{\operatorname{Re}(\bar u\partial_j u)}s,
 \qquad s v_j=\frac{\operatorname{Re}(\bar uD_ju)}s.
\end{equation}
Writing $q=|\nabla s|^2$, $f=s^2|\bm v|^2$, and $\chi=2s(v_y\partial_xs-v_x\partial_ys)$ gives $t=q+f$ and $d=q+f+\chi$. Pointwise discrepancies in these identities are below $10^{-12}$ when normalized by the maximum kinetic density. At exact zeros, only the amplitude--flow decomposition is omitted; the excluded kinetic weight is below $3.1\times10^{-12}$ of $T$. The direct defect and kinetic integrals include all grid points.

Table~\ref{supp:tab:fullspatial} gives the complete attractive sequence. The positive kinetic and amplitude-gradient densities have no peripheral enhancement comparable to the defect. With the re-converged zero-coupling field, the outer defect fraction grows monotonically along the whole sequence, while the kinetic and amplitude-gradient fractions stay near or below the $10\%$ mass share.
\begin{table}[!htbp]
\centering\small
\caption{Spatial response at $\beta=61$. Entries are percentages of each quantity outside $R_{0.9}$. A dagger marks $r_{\rm rel}>10^{-3}$. The last column compares regional nonsaturation, with $\ell_\Omega=(\int_\Omega d(\rvec)\,\dd\rvec)/(\int_\Omega t(\rvec)\,\dd\rvec)$.}\label{supp:tab:fullspatial}
\begin{tabular}{rrrrrr}
\toprule
$\kappa$ & Defect & Kinetic & Amplitude & Quartic & $\ell_{\rm out}/\ell_{\rm in}$\\
\midrule
\spatialrows
\bottomrule
\end{tabular}
\end{table}

\section*{S6. Variational comparison}
\begin{figure}[ht]
\centering
\includegraphics[width=.96\textwidth]{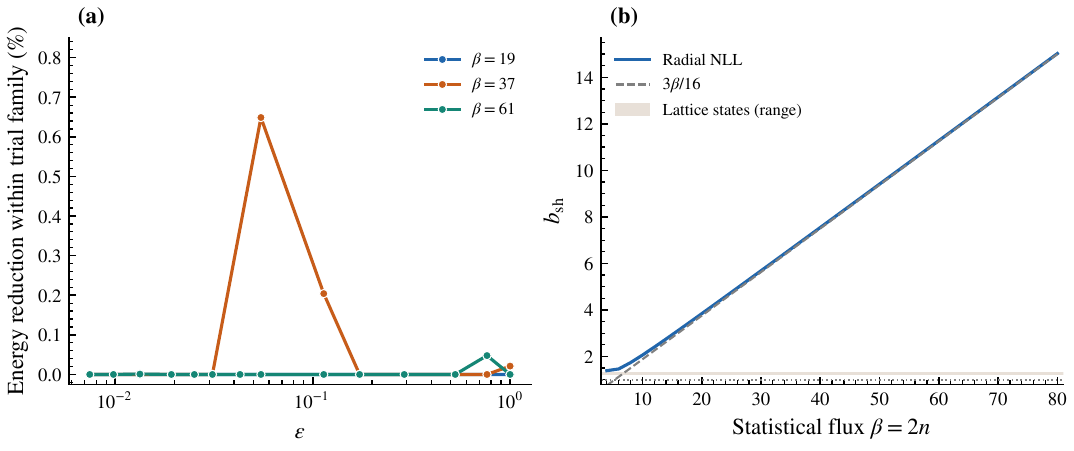}
\caption{Variational comparisons. (a) Energy reduction from selecting and dilating the best stored attractive shape. (b) Shape factor of the exact radial NLL at even flux and its $3\beta/16$ asymptote. The shaded band spans the odd-flux lattice values; it is a comparison of geometry across fluxes, not a matched-flux energy ordering.}\label{supp:fig:trialfamily}
\end{figure}

For a stored shape, dilation optimization gives $U_i(G)^2=4P_i(T_i+GI_i/2)$. The minimum over a finite family is concave by construction. Figure~\ref{supp:fig:trialfamily} compares this energy with the selected states and gives the radial NLL shape factor. Its positive energy at the boundary is a property of the trial family, not a determination of the exact trapped critical infimum. The numerical sequences characterize finite-resolution low-energy states; they do not establish a universal endpoint coefficient.

\section*{S7. Confirmation runs}
\begin{figure}[ht]
\centering
\includegraphics[width=.96\textwidth]{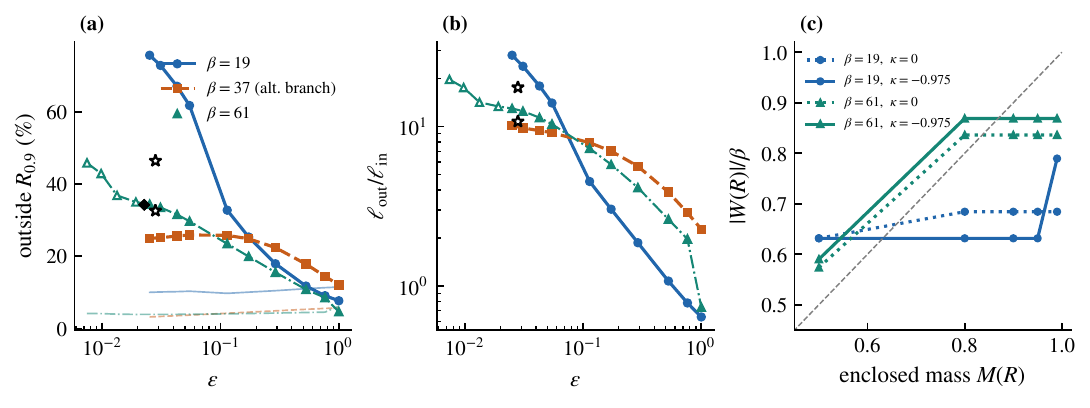}
\caption{Flux dependence and robustness. (a) Defect fraction outside $R_{0.9}$ (symbols) and kinetic fraction (thin lines) for $\beta=19$, the recomputed higher-energy $\beta=37$ branch, and $\beta=61$; open symbols mark states above the stationarity criterion, stars independent starts, and diamonds the grid and box audit. (b) Regional nonsaturation ratio. (c) Phase winding on mass-radius circles divided by $\beta$; the dashed diagonal is the enclosed statistical flux. The plateau of $|W|$ beyond $M\simeq0.8$ marks the vortex-free periphery.}\label{supp:fig:confirm}
\end{figure}

The script \texttt{anyon\_confirm\_kaggle.py} repeats the calculations with the same solver and adds the phase winding $W(R)$, evaluated by cubic interpolation of $u$ on circles that enclose fixed fractions of the mass. Table~\ref{supp:tab:confirm} lists the $\beta=61$, $\kappa=-0.975$ tests and Table~\ref{supp:tab:zero} the zero-coupling starts. Figure~\ref{supp:fig:confirm} compares the three fluxes.
\begin{table}[ht]
\centering\small
\caption{$\beta=61$, $\kappa=-0.975$: further optimization, grid and box refinement, and independent starts. Outer fraction and ratio refer to $R_{0.9}$; $N_v$ is the number of plaquette-detected vortices. A dagger marks $r_{\rm rel}>10^{-3}$.}\label{supp:tab:confirm}
\begin{tabular}{lrrrrrr}
\toprule
Test & $E$ & $\delta$ & $r_{\rm rel}$ & Outer defect (\%) & $\ell_{\rm out}/\ell_{\rm in}$ & $N_v$\\
\midrule
\confirmrows
\bottomrule
\end{tabular}
\end{table}
\begin{table}[ht]
\centering\small
\caption{$\beta=61$ at zero coupling: polished archived field and two fresh starts. The polished field is used in the paper. Columns as in Table~\ref{supp:tab:confirm}.}\label{supp:tab:zero}
\begin{tabular}{lrrrrr}
\toprule
Start & $E$ & $\delta$ & $r_{\rm rel}$ & Outer defect (\%) & $N_v$\\
\midrule
\confirmzerorows
\bottomrule
\end{tabular}
\end{table}

\section*{S8. Edge mechanism}
\begin{figure}[ht]
\centering
\includegraphics[width=.96\textwidth]{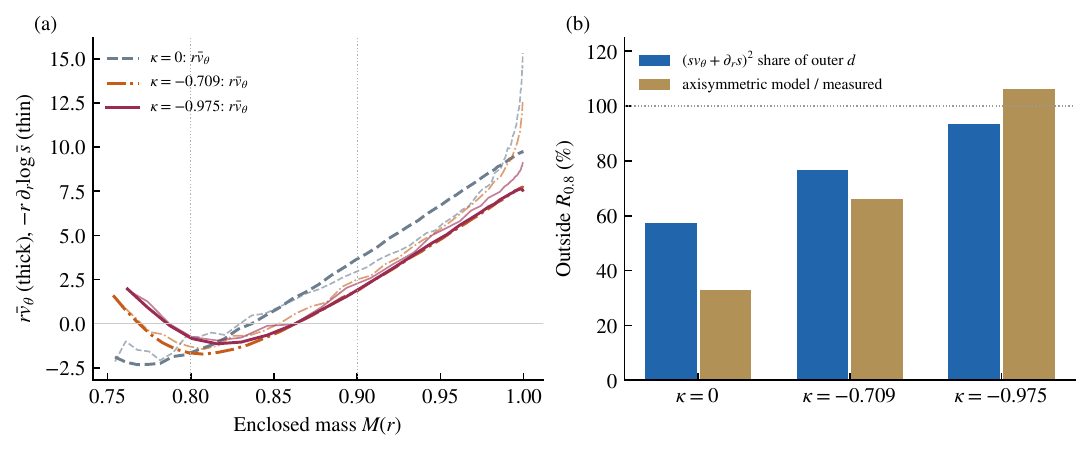}
\caption{Vortex-free edge at $\beta=61$. (a) Angular-averaged azimuthal flow $r\bar v_\theta$ (thick) and density decay rate $-r\partial_r\log\bar s$ (thin) versus enclosed mass; local self-duality requires equality. Dotted lines mark $M=0.8$ and $0.9$. (b) Share of the defect outside $R_{0.8}$ carried by $(sv_\theta+\partial_rs)^2$, and the ratio of the axisymmetric model defect to the directly integrated outer defect.}\label{supp:fig:edge}
\end{figure}

The script \texttt{edge\_mechanism.py} reconstructs, from the stored $\beta=61$ fields, the angular averages $\bar\rho(r)$ and $\bar J_\theta(r)$ in annuli of width $h$ and forms $\bar v_\theta=\bar J_\theta/\bar\rho$ and $\bar s=\bar\rho^{1/2}$. The circulation $r\bar v_\theta-\beta M(r)$ is constant to within $0.2$ outside $R_{0.9}$ and equals $-53.1$ at $\kappa=-0.975$ and $-51.2$ at zero coupling, reproducing the plaquette winding independently. The axisymmetric model defect is $2\pi\int_{R_{0.8}}^\infty r\bar\rho(\bar v_\theta+\partial_r\log\bar s)^2\,\dd r$. Figure~\ref{supp:fig:edge} compares the flow with the decay rate and the model with the directly integrated defect. In the closest, unaccepted state at $\kappa=-0.993$ the azimuthal share and model ratio are $78\%$ and $0.78$.
\end{document}